\documentclass{aa}  
\usepackage{graphicx}
\usepackage{txfonts}
\usepackage[hidelinks]{hyperref}
\usepackage{amsmath}
\usepackage{amssymb}    
\usepackage[T1]{fontenc}
\usepackage{ae,aecompl}
\usepackage{xcolor}
\usepackage{multirow}
\usepackage{tabularx}
\usepackage{threeparttable}
\usepackage[utf8]{inputenc}
\usepackage{natbib}
\usepackage{adjustbox}
\bibpunct{(}{)}{;}{a}{}{,} 

\usepackage{siunitx}

\begin{document}

   \title{Evidence of a non-conservative mass transfer in the ultra-compact X-ray source XB 1916-053}

 

\author{R. Iaria\inst{1}, A. Sanna\inst{2,3}, T. Di Salvo\inst{1,4}, A. F. Gambino\inst{1}, S. M. Mazzola\inst{1},  A. Riggio\inst{2,3}, A. Marino\inst{1,5,6},  L. Burderi\inst{2,3}}

  \institute{Dipartimento di Fisica e Chimica - Emilio Segrè, Universit\`a di Palermo, via Archirafi 36 - 90123 Palermo, Italy
        \and
         Dipartimento di Fisica, Universit\`a degli Studi di Cagliari, SP Monserrato-Sestu, KM 0.7, Monserrato, 09042 Italy
 \and
  INAF - Osservatorio Astronomico di Cagliari, via della Scienza 5, 09047 Selargius (CA), Italy   
  \and 
  INFN, Sezione di Cagliari, Cittadella Universitaria, 09042 Monserrato, CA, Italy 
  \and 
  Istituto Nazionale di Astrofisica, IASF Palermo, Via U. La Malfa 153, I-90146 Palermo, Italy
  \and
  IRAP, Universitè de Toulouse, CNRS, UPS, CNES, Toulouse, France  }
   
 
  \abstract
   {The dipping source XB 1916-053 is a compact binary system with an orbital period of 50 min harboring a neutron star. 
   It shows a positive and a negative superhump, which suggests the presence of a precessing elliptic accretion disk  tilted  with respect to the equatorial plane of the system. 
   The orbital ephemeris indicates a large orbital period derivative, $\dot{P}/P=1.53 \times 10^{-7}$ yr$^{-1}$, that can be explained assuming a high non-conservative mass transfer rate. Finally, the spectrum shows prominent absorption lines  indicating the presence of an ionized absorber along the line of sight.}
   {Using ten new  {\it Chandra} observations and one {\it Swift/XRT} observation, we are able to extend the baseline of the orbital ephemeris; this allows us to exclude some   models that explain the dip arrival times. The Chandra observations provide a good plasma diagnostic of the ionized absorber and allow us to determine whether  it is placed at the outer rim of the accretion disk or closer to the compact object. }
   {From the available observations we are able to obtain three new dip arrival times extending the  baseline of the orbital ephemeris from 37 to 40 years.  The {\it Chandra} spectra are fitted  adopting  a Comptonized continuum. To fit the absorption lines we adopt the   {\sc zxipcf} component
   obtaining information on the ionization parameter and the equivalent hydrogen column density of the ionized absorber. }
   {From the analysis of the dip arrival times 
   we confirm an orbital period derivative of  $\dot{P}=1.46(3) \times 10^{-11}$ s s$^{-1}$. Furthermore, 
   the unabsorbed 0.1-100 keV luminosity observed from the {\it Chandra}   spectra show a variation
   between $3 \times 10^{36}$ and $1.4 \times 10^{37}$ erg s$^{-1}$.  We show that the  $\dot{P}$ value and the luminosity values are compatible with   neutron star masses higher than  1.4 M$_{\odot}$ with a mass accretion rate lower than 10\% of the mass transfer rate.  We show that  the mass ratio $q=m_2/m_1$ of 0.048 explains the apsidal precession period of 3.9 d and the nodal precession period of 4.86 d deduced from the superhump and infrahump detected period.  
    
   The observed absorption lines are associated with the presence of \ion{Ne}{x}, \ion{Mg}{xii}, \ion{Si}{xiv}, \ion{S}{xvi,} and \ion{Fe}{xxvi} ions. We observe a redshift in the absorption lines between $1.1 \times 10^{-3}$ and $1.3 \times 10^{-3}$. By interpreting it as gravitational redshift, as recently discussed in the literature,  we  find that   the ionized absorber is placed at   a distance of  
   $10^8$ cm from the  neutron star with a mass of 1.4 M$_{\odot}$ and  has a hydrogen atom   density greater than  $10^{15}$ cm$^{-3}$. Instead, the absorber is 
   more distant and  could be placed at the outer rim of the accretion disk  ($1.7 \times 10^{10}$ cm) during the dip activity.}
  {We show that the mass ratio of the source is 0.048; this value is obtained   from the nodal precession period of the disk and from  the apsidal precession period taking into account the pressure term due to the spiral wave present in the disk.  From our analysis 
  we estimate  a pitch angle of the spiral wave smaller than 30$^{\circ}$, in agreement with the values observed in several cataclysmic variables. We show that the outer radius of the disk is truncated at the  radius in which a 3:1 resonance occurs, which is $1.7 \times 10^{10}$ cm for a neutron star mass of 1.4 M$_{\odot}$. The large orbital period derivative is likely due to a high non-conservative mass transfer with a mass transfer rate of $10^{-8}$ M$_{\odot}$ yr$^{-1}$.  
  The variation in observed luminosity could be explained assuming that the ejection point from which the matter leaves the system moves close to the inner Lagrangian point.}
 
  \authorrunning{R. Iaria et al.}

  \titlerunning{Evidence of a  non-conservative mass transfer in  XB 1916-053}
  
  

   \maketitle

\section{Introduction}

XB 1916-053 is a low-mass X-ray binary system (LMXB) showing dips and type I X-ray bursts. The source was the first LMXB in which periodic absorption dips were detected  
\citep{Walter_82, White_82}, the analysis of which allowed   an orbital period 
 of 50 min and its compact nature to be estimated.  
 
 The optical counterpart of the source (a V=21 star) was discovered by
 \cite{Grindlay_88}. 
 \cite{Swank_84} discussed that the companion star (CS) is not hydrogen exhausted by analyzing the thermonuclear flash models of X-ray bursts, while \cite{Pac_81}
  showed that X-ray binary systems with orbital periods shorter than 81 min cannot contain hydrogen-rich CS. 
   \cite{Galloway_08}, by studying two consecutive 
    type I X-ray bursts temporally separated by 6.3 hours, 
   estimated that the CS contains a 20\% fraction of hydrogen under the hypothesis of a  NS mass of 1.4  M$_{\odot}$. 
 The same authors  estimated  a distance to the source of  8.9 kpc.

The optical light curve of the source shows a periodicity   of 
$3027.4 \pm 0.4$ \citep{Grindlay_88}. \cite{Callanan_95}
found that the optical period remained stable over a baseline of seven years, and refined the  
    period estimate to  $3027.551\pm0.004$ s. 
The most recent X-ray orbital ephemeris of the source  gives an  orbital period of 
$3000.6496\pm 0.0008$ s and an orbital period derivative of 
$1.44(6) \times 10^{-11}$ s s$^{-1}$ \citep{iaria_15}.
The 1\% discrepancy between the optical and X-ray periods was explained by \cite{Grindlay_88}, who invoked the presence of a third body with a period of 2.5 d and a retrograde orbit.  This scenario predicts that the optical period reflect  the real period of the binary system, while 
the observed X-ray period is modified by an increase in mass transfer influenced by the presence of the third body. 

An alternative scenario suggests that the real period of XB 1916-053 is actually the observed X-ray period. The  SU Ursae Majoris (SU UMa)   superhump scenario was initially adopted by \citet{White_89}.
The superhump phenomenon is the appearance of a periodic or  quasi-periodic modulation in the light curves of the SU UMa class of dwarf novae while showing   superoutburst activity.   The period of the superhumps is  longer than the orbital period of the system in which it is observed. The commonly accepted interpretation of the phenomenon ties the  longer modulations   to the  beat between the binary period and the apsidal precession period of the elliptical accretion disk.  

Hydrodynamic simulations show that   accretion disks in   cataclysmic variable (CV) systems with a low-mass secondary are tidally unstable with a high probability of forming 
 a precessing eccentric disk \citep{whur_88}. \cite{Hirose_90} 
found the dependence of the dynamical term from the apsidal precession period of the disk.   \cite{Chou_01}   analyzed  optical and X-ray data estimating the 
 superhump period  and  the associated apsidal precession period of $3.9087 \pm 0.0008$ days. Adopting the SU SMa scenario and the relation discussed by  \cite{Hirose_90}, 
\cite{Chou_01}  inferred a mass ratio $q=M_2/M_1$ of 0.022, where $M_1$ and   $M_2$  are the NS and CS mass, respectively. 

\cite{Retter_02} detected a further  X-ray periodicity at 2979 s in the RXTE light curves of XB 1916-053.
A similar periodicity had already been observed by \cite{Smale_89} analyzing {\it Ginga} data. 
\cite{Retter_02}  interpreted the 2979 s period invoking the presence of a negative superhump (also called an infrahump) where  the X-ray period is the orbital period of the system  that beats with 
the nodal precession period of 4.86 days. This means that the disk is tilted with respect to the equatorial  plane. 
Finally, \cite{Hu_08},  by adopting the relation between the nodal angular frequency and the mass ratio $q$ proposed by \cite{Larwood_96} and assuming a nodal precession period of 4.86 days,  inferred $q \simeq 0.045$.

\cite{iaria_15} used a baseline of 37 years to
infer the accurate orbital ephemeris of the source 
from the dip arrival times.  Although they proposed several models to describes the dip arrival times, the authors focused their attention on the ephemeris containing a quadratic term plus a long sinusoidal modulation. The quadratic term 
implies the presence of an orbital period derivative of 
$1.44(6) \times 10^{-11}$ s s$^{-1}$, while the long modulation of 25 years was explained with the presence of third body with mass of 0.055 M$_{\odot}$ orbiting around the binary system. \cite{iaria_15}  showed that 
the large orbital period derivative can be explained only assuming a high non-conservative mass transfer rate with only   8\% of the mass transfer rate accreting onto the NS.

The spectrum of XB 1916-053 shows  absorption lines superimposed on the continuum emission. The  \ion{Fe}{xxv} and \ion{Fe}{xxvi}  absorption lines were detected for the first time by \cite{Boirin_04} from the analysis of {\it XMM-Newton} data. 
\cite{Iaria_06}, using {\it Chandra} data, detected   prominent  absorption lines in the spectrum associated with
the presence of \ion{Ne}{x}, \ion{Mg}{xii}, \ion{Si}{xiv},
\ion{S}{xvi,} and   \ion{Fe}{xxvi} ions. From the 
plasma diagnostic the authors proposed that the lines should be produced at the outer rim of the accretion disk. 
 Analyzing the same data, \cite{Juett_06} suggested 
 a thickness for  the X-ray absorber of    $<3.2\times 10^9$ cm, assuming that it is located at the outer edge of the accretion disk. 
 Finally, analyzing {\it Suzaku}   data, \cite{Gambino_19}    set an upper limit of $<1 \times 10^{10 }$  cm  
 on the distance of the the ionized absorber from the NS,
placing it at the innermost region of the accretion disk.

In this work we update the orbital ephemeris of XB 1916-053 
by expanding the available baseline to 40 years using ten {\it Chandra} observations and one {\it Swift/XRT} observation.  We revise 
the estimation of the NS mass suggesting that    a   mass of 1.4 M$_{\odot}$ can also explain the observed orbital period derivative and   luminosities.  
From theoretical considerations we discuss that a mass ratio of 0.048 can explain both the apsidal and the nodal precession period of the accretion disk taking into account the presence of a spiral wave in the accretion disk with a pitch angle lower than 30$^{\circ}$. 
From the spectroscopic analysis we study the absorption lines detected in the spectrum during the persistent emission at different source luminosities and during the dips.

We note that the same {\it Chandra} data have recently  been analyzed  by \cite{Trueba_20}. The authors observed a redshift in the absorption lines that can be interpreted as a gravitational redshift.  We reached similar results by assuming the same scenario, but adopting a different analysis. 

\section{Observations} 
\begin{table} 
\centering
\begin{threeparttable}
\caption{List of  observations}
\begin{tabular}{l l  c c c }
\hline
\hline
ObsID. & Start Time (UT)   &  Exposure & Type I  \\
       &           & Time (ks) & bursts & \\
       \hline
20171 & 2018 June 11 20:11:59 & 22 &  1  \\ 
21103 & 2018 June 12 09:08:20 & 29 & 1 \\
21104 & 2018 June 13 17:03:34 & 23 &  1 \\
21105 & 2018 June 15 23:35:18 & 21.7 &  1 \\
20172 & 2018 July 31 08:02:08 & 30.5 & no \\
21662 & 2018 August 01 06:48:53 & 29.5 &  1 \\
21663 & 2018 August 02 19:47:53 & 30.5 & no  \\
21664 & 2018 August 05 05:09:45 & 22.4 & no\\
21106 & 2018 August 06 03:26:33 & 21 & no\\
21666 & 2018 August 06 18:04:31 & 19 & no\\ 
4584{\textit{$\rm ^a$}}   & 2004 August 07 02:34:45 &  46  & 2 \\  
 \hline
\hline
\end{tabular}
\begin{tablenotes}
\item[a] Observation 4584 was already analyzed by \cite{Iaria_06}.
\end{tablenotes}
\label{tab:1}
 \end{threeparttable}
\end{table}  
\begin{figure*}
\centering
\includegraphics[scale=.50]{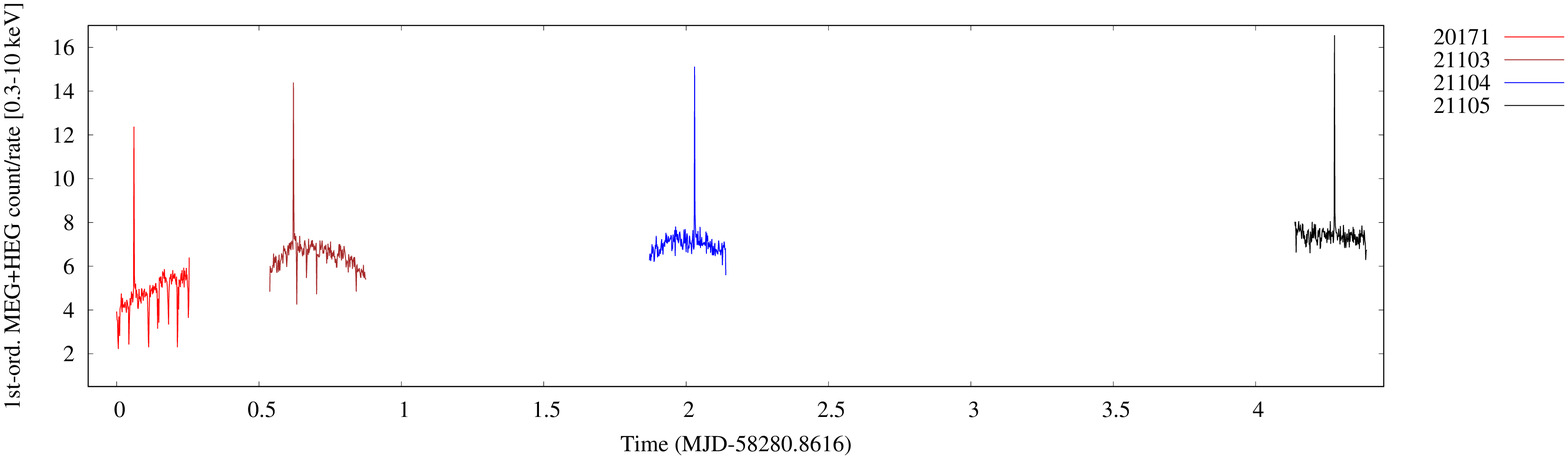}\\
\includegraphics[scale=.50]{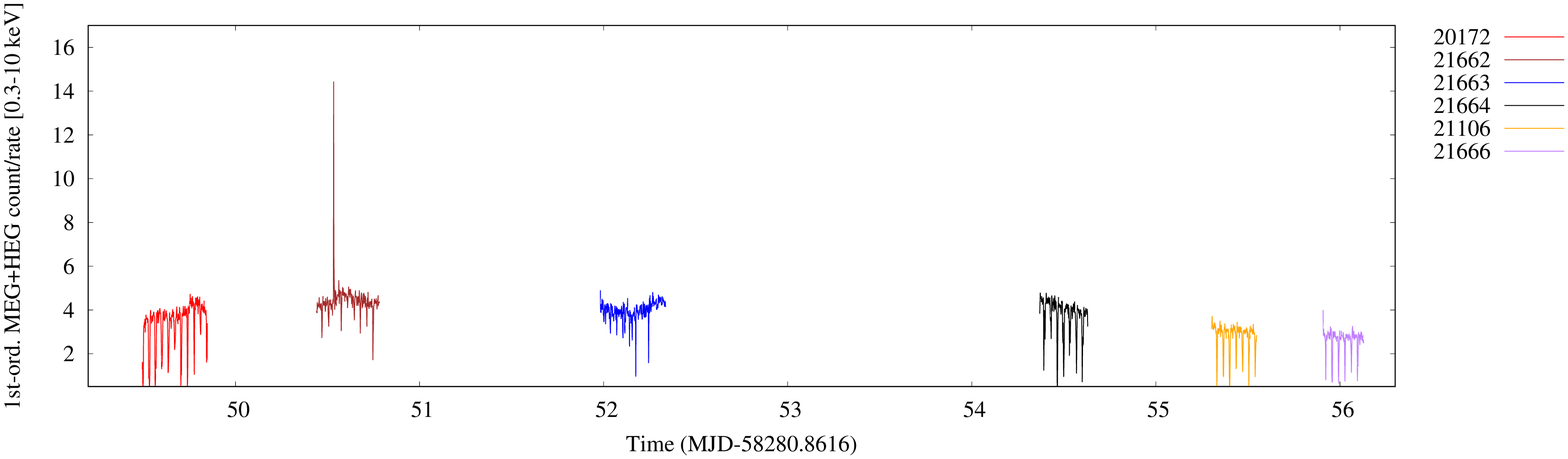}
\caption{First-order MEG+HEG light curves of the observations in the 0.3-10 keV energy range.  The bin time is 128 s.}
\label{figure:1}
\end{figure*}

We analyzed ten new observations of  XB 1916-053 collected by the \textit{Chandra} observatory in 2018 and the Swift/XRT observation (ObsID. 87248021) taken on 2017 July 30; furthermore,  we  reanalyzed the Chandra observation taken in 2004 August 7 \citep[see][]{Iaria_06}.

The new {\it Chandra} observations  were performed  from 2018 June 11 to August 6 using the onboard High Energy Transmission
Grating Spectrometer (HETGS) in timed graded mode;  the detailed list of the observations is shown in Table  \ref{tab:1}.  HETGS consists of two types of transmission gratings, the medium energy grating (MEG) and the high-energy grating (HEG). The HETGS provides high-resolution spectroscopy from 1.2 to 31 $\AA$  (0.4-10 keV) with a peak spectral resolution of $\lambda/\Delta\lambda \sim$1000 at 12 $\AA$ for  first-order HEG. The dispersed spectra were recorded with an array of six charge-coupled devices (CCDs) that are part of the Advanced CCD Imaging Spectrometer-S.

We reprocessed the data using the FTOOLS ver. 6.26.1 and the CIAO ver. 4.11 packages.
The brightness of the source required additional efforts to mitigate  pileup issues. A 512 row subarray (the first row = 1) was applied during the observations, reducing the CCD frame time to 1.7 s. We ignored the zeroth-order events in our  analysis focusing on  the first-order HEG and MEG spectra.
We reprocessed the data adopting a scale factor (width$\_$factor$\_$hetg) that multiplies the approximate one sigma width of the HEG/MEG mask in the cross-dispersion direction of 15, instead of the default value of 35, in order to avoid  that HEG and MEG  overlap  in the Fe K region of the spectrum.  We obtained HEG and MEG region widths of 33 and 44 pixels, respectively. 

We applied  barycentric correction to the events adopting the source coordinated obtained by \cite{Iaria_06} and extracted the light curve for each observation retaining 
only the  first-order MEG and HEG data in the 0.3-10 keV energy range. We show the light curves of the ten observations in Fig. \ref{figure:1}. During the persistent emission, the observations 
taken in 2018 June (ObsID. 20171, 21103, 21104, 21105) have  a higher count rate than those taken in July-August (ObsID. 20172, 21662, 21663, 21664, 21106, 21666). Observations 
21103, 21104, and 21105 show a count rate close to 7 c s$^{-1}$ (20171 
shows a count rate of 5 c s$^{-1}$);    observations 20172, 21662, 21663, and 21664 show a count rate of  4.5 c s$^{-1}$;  and observations 21106 and 21666 show a count rate of 3  c s$^{-1}$. We observed five type I X-ray bursts, four during the high-flux state and one when the flux of the source dropped below 5 c s$^{-1}$. Furthermore, the light curves corresponding to  observations 21104 and 21105 (when XB 1916-056 showed a high flux) do not show dips, while the dipping activity is intense at    lower fluxes. 
A similar behavior  was observed during a long Suzaku observation  of the source \citep[see][]{Gambino_19}.

XB 1916-053 was monitored by the {\it X-Ray Telescope}
\citep[XRT;][]{Burrows_05}
 on board the Swift Observatory \citep{Ger_04}. We analyzed the Swift/XRT observation 87248021  taken on 2017 July 30 
10:54:17, with elapsed and exposure times of
87 ks and 8.9 ks, respectively. The observation was  performed in photon counting (PC) mode. The data were locally reprocessed by the UK Swift Science Data Center 
using HEASOFT v6.26.1. We applied barycentric correction to the events using the ftool {\tt barycorr}\footnote{https://www.swift.ac.uk/analysis/xrt/barycorr.php}. Using XSELECT we extracted the source events in the 1-4 keV energy range adopting a circular region of 100$\arcsec$; the background events in the same energy range were extracted using a same size circular region 
free from the source. Finally, we applied the exposure-correction to  the source and background-light curves and extracted the background-subtracted light curve using the ftools {\tt xrtlccorr} and {\tt lcmath}\footnote{https://www.swift.ac.uk/analysis/xrt/lccorr.php}.

\section{Updated orbital ephemeris of XB 1916-053}
First, we filtered the light curves to exclude    the type I X-ray bursts; then we  selected the events in the 1-4 keV energy range  to estimate the dip arrival times. 
 We estimated one dip arrival time from the Chandra observations taken in June, one from those taken in July-August, and  one from the XRT observation. 
 
 The light curves were folded adopting  $P_0=$3000.6511 s as the orbital period.
 For the first dip arrival time we folded the light curve obtained from  observation 20171, where the dips are the deepest, adopting a folding time 
$T_{fold}=$58281 MJD; to obtain the second  we folded together the light curves of  observations 20172, 21664, 21106, and 21666, adopting   
$T_{fold}=$58333.65 MJD. Finally, the XRT light curve was folded assuming   $T_{fold}=$57965 MJD.

 We fitted the dips with a simple model consisting of a step-and-ramp function, where the count rates before,
during, and after the dip are constant and the intensity changes
linearly during the dip transitions \citep[see][for a  description]{iaria_15}. We show   
the X-ray dip arrival times  in Table \ref{tab:2}. \begin{table} 
\centering
\begin{threeparttable}
\caption{Journal of the estimated dip arrival times}
\begin{tabular}{l@{\hspace{2pt}}c@{\hspace{\tabcolsep}}c@{\hspace{\tabcolsep}}l@{\hspace{\tabcolsep}}}
\hline
\hline
Dip Time & Cycle  & Delay (s) & Observation\\
 (MJD;TDB)  &           &  \\
 \hline
 57965.0004(2)& 225800& $1062 \pm 14$ & Swift/XRT\\
58281.0075(2) & 234899 & $1155 \pm 21$ & 1st ord. HEG+MEG\\
58333.65781(6) & 236415 & $ 1151 \pm 8$& 1st ord. HEG+MEG\\ 
 \hline
\hline
\end{tabular}
 Epoch of reference 50123.00873 MJD, orbital period 3000.6511 s.   
\label{tab:2}
 \end{threeparttable}
\end{table}  
To obtain the delays with respect to a constant period reference, we used the period   $P_0= 3000.6511$ s and reference epoch   $T_0= 50123.00873$ MJD. Hence, we added the three dip arrival times to the 27 values reported by \cite{iaria_15}.  

\begin{figure} 
\centering
\includegraphics[scale=.5]{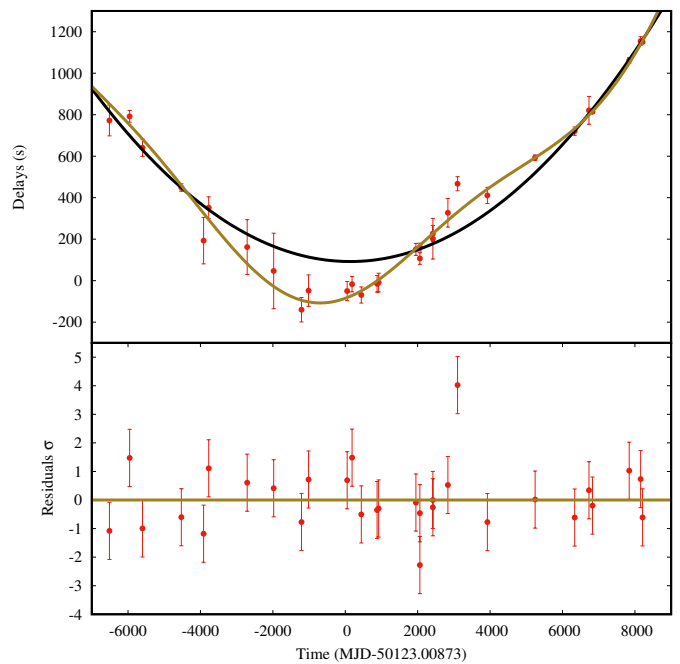}
\caption{Top panel:  Delays vs time (red points). The black curve is the best fit of the LQ function. The brown curve is the best fit of the LQS function. Bottom panel: Residuals in units of $\sigma$ vs time, corresponding to the LQS function}
\label{figure:3}
\end{figure}

We initially fitted the  delays with a quadratic function 
$
y(t) = a+b t+ c t^2$,
where $t$ is the time in days with respect to $T_0$, $a=\Delta T_0$ is
the correction to $T_0$ in units of seconds, $b=\Delta P/P_0$ in units
of s d$^{-1}$  with $\Delta P$ the correction to the orbital period, and
 $c= 1/2 \;\dot{P}/P_0$ in units of s d$^{-2}$ with
$\dot{P}$  is the orbital period derivative.  The quadratic form
does not fit the data well, returning a  $\chi^2({\rm {d.o.f.}})$ of
182(27). We show the delays  (red points)  and the quadratic curve (black curve) 
 in the top panel of  Fig. \ref{figure:3}.
The best-fit values of the parameters are shown in the second column of  
Table \ref{table:fit_result}. 
\begin{table}
\setlength{\tabcolsep}{3pt} 
  \caption{Best-fit values of the functions adopted to fit the 
delays.\label{table:fit_result}}
\begin{center}
\begin{tabular}{l@{\hspace{2pt}}c@{\hspace{\tabcolsep}}c@{\hspace{\tabcolsep}}c@{\hspace{\tabcolsep}}}          
\hline                                             
\hline  
Parameters & LQ   & LS & LQS \\  
\hline                                             
$a$   (s)      & 
$93 \pm 22$  &    
$5495\pm300$ & 
$14 \pm 14$   \\

$b$   ($\times 10^{-3}$ s d$^{-1}$)    & 
$-4 \pm 4$ & 
$-408\pm144$    & 
$5 \pm 3$ \\

$c$   ($\times 10^{-5}$ s d$^{-2}$)&
$1.64 \pm 0.07$ & 
        --   &
$1.82 \pm 0.04$ \\

$A$   (s)      &  
--  & 
$7343\pm1380$    & 
$130 \pm 14$   \\

$t_\phi$ (d)      & 
--   &  
$9826\pm1795$   & 
$1262 \pm      133$ \\

$P_{mod}$  (d)  & 
-- & 
73574 (fixed)   & 
$9099 \pm 302$  \\

$\chi^2$(d.o.f.) &
182(27)& 
132(26) &
37(24)  \\
\hline   
\hline                            
\end{tabular}
\end{center} {\small \sc Note} \footnotesize--- The reported errors
are at 68\% confidence level.  The best-fit parameters of the delays
are obtained using the functions LQ (column 2), LS (column 3), and  
LQS  (column 4).
\end{table}

To improve the   modeling of the delays we added 
 a cubic
term to the previous parabolic function,  
$
y(t) = a +bt+ct^2+dt^3$,
where  $d$ is
defined as $\ddot{P}/(6 P_0)$ and  $\ddot{P}$ indicates the
temporal second  derivative of the orbital period. Fitting with this
cubic function, we obtained a $\chi^2({\rm {d.o.f.}})$ of 129(26) with an F-test probability of chance
improvement of only $3 \times 10^{-3}$ with respect to the
quadratic form (hereafter LQ function).   
 We also tried to fit the delays using a linear 
plus a sinusoidal function \citep[the LS function in][]{iaria_15},  
$y(t) = a+b t+ A \sin\left[\frac{2 \pi}{P_{mod}}(t-t_\phi)\right],$ where $a$ and $b$ are defined as above, while $A$ and $P_{mod}$ are the amplitude  in seconds and the period in days  of the sinusoidal function, respectively.  Finally, $t_\phi$ is the time in days 
referred to $T_0$ at which 
the sinusoidal function is null. Since the best-fit value of $P_{mod}$ is larger then the available time span, we fixed it   to the best-fit value   $P_{mod}=73574$ d.
We found a $\chi^2({\rm {d.o.f.}})$ of 132(26); the best-fit parameters are shown in the third column of Table \ref{table:fit_result}. 

We added a quadratic term to the LS function to take  into account
the possible presence of an orbital period derivative  
 $y(t) = a+b t+ c t^2 + A \sin\left[\frac{2 \pi}{P_{mod}}(t-t_\phi)\right]$  (hereafter LQS function).
 We obtained a value of $\chi^2({\rm {d.o.f.}})$ of
37(24) and an F-test probability of chance improvement 
with respect to the LQ function of $1.8 \times 10^{-8}$.
We show the best-fit curve of the LQS model (brown curve)
and its relative residuals in units of sigma in the top and bottom panels  of Fig. \ref{figure:3}, respectively. 
 The best-fit values are shown in the forth column of Table \ref{table:fit_result}. 
The obtained best-fit values are compatible with the 
previous ones reported by \cite{iaria_15}.

 The corresponding  LQS ephemeris  is
\begin{equation}
\label{linear_quad_sin_eph}\begin{split}
T_{dip}(N) = {\rm MJD(TDB)}\; 50\,123.0089(2) + \frac{3\,000.65129(8)}{86\,400}  N +\\
+2.54(6) \times 10^{-13} N^2 + \frac{A }{86\,400} \sin\left[\frac{2 \pi}{P_{mod}}\left(t -t_{\phi}\right)\right], 
\end{split}\end{equation}
with $P_{mod} = 9099  \pm 302 $ d ($24.9\pm0.8$ yr), $t_{\phi} = 1262 \pm 133$ d, and $A=130 \pm 14$ s. 
The corresponding orbital period derivative is $\dot{P} = 1.46(3) \times  10^{-11}$ s s$^{-1}$.
  Our analysis  suggests  that a quadratic or a quadratic plus a cubic
term do not   adequately fit the delays, as already shown by \cite{iaria_15}.

\begin{figure}
\centering
\includegraphics[scale=.52]{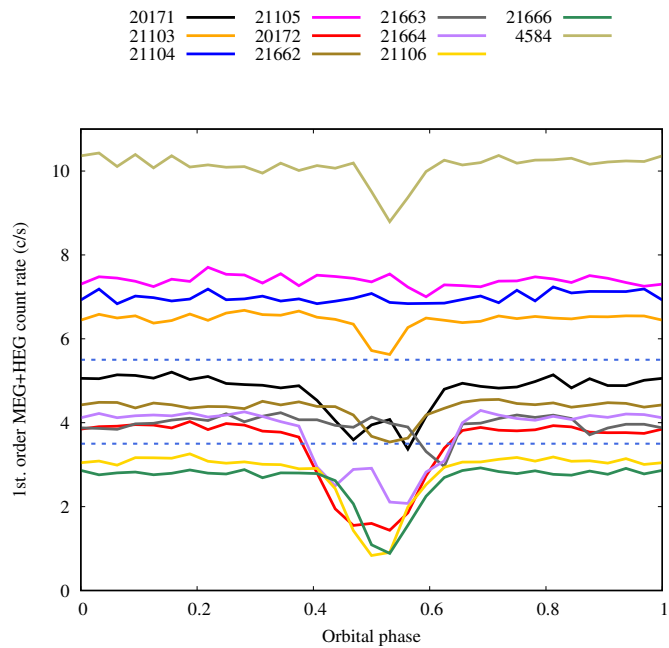}
\caption{First-order MEG+HEG  folded light curves in the 0.3-10 keV energy range adopting the  LQS ephemeris. An ad hoc phase shift of 0.5 is imposed. The label in the plot shows the correspondence 
between   colors and  observations. The blue dashed horizontal lines indicate the arbitrary count rate threshold of 3.5 c s$^{-1}$ and 5.5 c s$^{-1}$ adopted to select the four sets of observations (see text).} 
\label{figure:5}
\end{figure}

We  folded the 0.3-10 keV first-order MEG+HEG  light curves 
obtained from the observations  (type I X-ray bursts excluded), adopting the orbital period inferred by the  LQS ephemeris and changing the  reference time so that the dip falls at the orbital phase 0.5.  The epoch-folded light curves are shown   
in Fig. \ref{figure:5}.

 We distinguish four different sets of epoch-folded light curves at different count rates: the first set (hereafter set A) associated with  observation 4584  has a persistent count rate higher than 10 c s$^{-1}$ (gold curve in Fig. \ref{figure:5});  the folded  curves of the second set (hereafter set B), 
 have a persistent count rate between  6 and 8 c s$^{-1}$ and are associated with observations 21103, 21104, and 21105  (orange, blue, and magenta curve in  Fig. \ref{figure:5}); the third set of curves (hereafter set C) has a   persistent count rate between 3.5 and 5.5 c s$^{-1}$  and is associated with observations 20171, 20172, 21662, 21663, and 21664 (black, red, brown, gray, and purple curve in Fig. \ref{figure:5});   
 the fourth set (hereafter set D) shows   a count rate close to 3.5 c s$^{-1}$ outside the dip associated with observations  21106 and 21666 (yellow and green curve in Fig. \ref{figure:5}).

 \begin{table}
\setlength{\tabcolsep}{3pt} 
  \caption{List of the combined spectra\label{table:4}}
\begin{center}
\scriptsize
\begin{tabular}{l@{\hspace{2pt}}c@{\hspace{\tabcolsep}}c@{\hspace{\tabcolsep}}c@{\hspace{\tabcolsep}}}          
\hline                                             
\hline  
Combined first-order & Exp. time & ObsID.   & Phase interval\\  
HEG+MEG  spectrum   &  (ks)  &    & excluded \\                                                  
\hline                                             
spectrum A & 38.8 & 4584 & 0.45-0.60 \\ \\                               
\multirow{3}{*}{spectrum B}   &  \multirow{3}{*}{67.3}       & 21104    &      \\
            &          & 21105    &      \\
            &          & 21103    & 0.45-0.60 \\\\
            
\multirow{5}{*}{spectrum C}& \multirow{5}{*}{104.2}      & 20171    & 0.40-0.65 \\  
            &          & 21662    & 0.45-0.60 \\  
            &          & 21663    & 0.55-0.66 \\
            &          & 21664    & 0.40-0.70 \\
            &          & 20172    & 0.40-0.65 \\\\
            
\multirow{2}{*}{spectrum D}    &  \multirow{2}{*}{31.4}    & 21106    & 0.40-0.60 \\  
            &          & 21666    & 0.40-0.60 \\  
\hline    
\hline   
\end{tabular}
\label{tab:excluded_phases}
\end{center} 
\end{table}


\begin{table*} 
\centering
\begin{threeparttable}
\caption{Best-fit values of {\tt Model 1} and {\tt Model 2}}
\scriptsize
\begin{tabular}{l@{\hspace{2pt}}l@{\hspace{\tabcolsep}}c@{\hspace{\tabcolsep}}c@{\hspace{\tabcolsep}}c@{\hspace{\tabcolsep}}c@{\hspace{\tabcolsep}}c@{\hspace{\tabcolsep}}c@{\hspace{\tabcolsep}}c@{\hspace{\tabcolsep}}c@{\hspace{\tabcolsep}}}
\hline
\hline
Model & Component &  \multicolumn{4}{c}{{\tt Model 1{\textit{$\rm ^a$}} }} & 
 \multicolumn{4}{c}{{\tt Model 2{\textit{$\rm ^b$}}}}\\
 &           & A & B & C &D & A & B & C & D\\

\hline

{\sc Constant} &C$_{\rm HEG}$  &  

$0.972\pm0.006 $ & 
$0.970 \pm 0.006$ & 
$0.966 \pm 0.006$ & 
$0.956 \pm 0.015$ &

$0.972\pm0.006 $ &  
$0.969 \pm 0.006$  &
 $0.966 \pm 0.006$  &
$0.956 \pm 0.015$ \\\\
               
 {\sc Edge}  & E (keV)  & 

\multicolumn{4}{c}{--} & 
\multicolumn{4}{c}{0.871 (fixed)} \\ 
 
  & $\tau$ & 
\multicolumn{4}{c}{--} &   
\multicolumn{4}{c}{$0.14\pm0.05$} \\ \\ 

{\sc TBabs} & N$_{\rm H}${\textit{$\rm ^c$}}  &    
$0.56 \pm 0.07$   & 
$0.63\pm0.08 $&  
\multicolumn{2}{c}{$0.53 \pm 0.09$} & 

$0.55  \pm 0.04$   &
$0.60\pm0.05$&
\multicolumn{2}{c}{$0.50\pm0.05$} \\\\            

{\sc partcov} & $f_{\rm cabs}${\textit{$\rm ^d$}} & 
\multicolumn{4}{c}{--} & 
\multicolumn{4}{c}{$>0.86$}\\

{\sc Cabs}  & N$_{{\rm H}_{\rm cabs}}${\textit{$\rm ^e$}}  & 

\multicolumn{4}{c}{--} & 
$19^{+7}_{-3}$  &
$14^{+5}_{-2}$ &
$17^{+6}_{-3} $&
$10\pm7$ \\\\

{\sc zxipcf}   & N$_{{\rm H}_{\xi}}${\textit{$\rm ^c$}}  & 
\multicolumn{4}{c}{--} & 
$19^{+7}_{-3}$  &
$14^{+5}_{-2}$ &
$17^{+6}_{-3} $&
$10\pm7$ \\\\

& log($\xi$) & 
\multicolumn{4}{c}{--} & 
\multicolumn{4}{c}{$4.33\pm0.02$}\\

& $f$ & 
\multicolumn{4}{c}{--} & 
\multicolumn{4}{c}{$>0.86$}\\

& $z$ ($\times 10^{-3}$)& 
\multicolumn{4}{c}{--} & 
$1.3\pm0.2$  &
$1.1\pm0.2 $&
$1.1 \pm 0.2$&
$1.0^{+1.3}_{-1.6}$\\\\

{\sc nthcomp}     &    $\Gamma$ &  
$1.613\pm0.012$  & 
$1.721\pm0.015 $   & 
$1.841\pm 0.015 $ & 
$1.80 ^{+0.02}_{-0.03}$ &

$1.620\pm0.009$ &
$1.730\pm0.013 $&
$1.848\pm0.014 $&
$1.803^{+0.014}_{-0.010} $\\       
        
     & kT$_{bb}$ (keV)  &     
 \multicolumn{4}{c}{$0.17\pm 0.05$}  & 
 \multicolumn{4}{c}{$0.16\pm0.04 $} \\

     & kT$_{e}$ (keV)  &     
 $2.9\pm 0.3$ & 
 $2.6 \pm 0.2$ & 
 $3.8^{+1.4}_{-0.9}$& 
 $3.3^{+4.9}_{-0.6}$&
 
 $3.2^{+0.5}_{-0.3}$&
 $2.6\pm0.3$&
 $>3.6$&
 $>2.8$\\          
                            
 & Norm ($\times 10^{-2}$) &     
 $11.7\pm 1.3$ & 
 $10.9 \pm 1.3$ & 
 $7.2\pm1.0$& 
 $4.8\pm0.7$&
 
 $14.3\pm1.0$&
 $12.8\pm1.0$&
 $8.7\pm0.8$&
 $5.4\pm0.5$\\\\ 
 
 & $\chi^2/dof$ &    
 \multicolumn{4}{c}{4957/4553} & 
 \multicolumn{4}{c}{3944/4542}\\

\hline
\hline
\end{tabular}
      \begin{tablenotes}
\item[] The associated errors are at 90\% confidence level.     
 \item[a] Model 1 = {\sc Const*TBabs*nthcomp}.
\item[b] Model 2 =  {\sc Const*TBabs*(partcov*Cabs)*zxipcf*nthcomp}.
\item[c] \textrm{Equivalent  hydrogen column density   in units of 10$^{22}$ atoms cm$^{-2}$}.
\item[d] The value of the parameter is tied to the value of $f$. 
\item[e] The value of the parameter is tied to the value of  N$_{{\rm H}_{\xi}}$. 
 \end{tablenotes}
\label{tab:mod1mod2}
 \end{threeparttable}
\end{table*}  
\section{Spectral analysis}
We extracted the combined spectrum from  each distinguished set, calling the spectra  A, B, C, and D,  using the CIAO script {\tt combine$\_$grating$\_$spectra} and 
 combining the corresponding first-order MEG  and HEG spectra. 
  Since our aim is the spectral analysis out of the dips, we excluded 
 the orbital phases in which the dip falls when it is present. The excluded phases are shown in Table \ref{table:4}. 
 Spectrum B was obtained  by   excluding the phases between 0.45 and 0.6 from observation 21103; the other two observations do not show dipping activity, and we included all the phases.  
 We rebinned the resulting first-order MEG and HEG spectra of  selections A, B, C, and D to have at least 250 counts  per energy channel. 
 Such a large rebinning could affect the spectroscopic study of narrow discrete features because it risks losing energy resolution. However, we  tried to fit these spectra using a rebinning so as to have 25 counts per energy channel, verifying that we obtained results that are compatible with those reported in the following.  
For the analysis we adopted the energy ranges   0.7-7 keV and 0.9-10 keV   for the first-order MEG and HEG spectrum, respectively.

To fit the spectra we used XSPEC v12.10.1p; we adopted the cosmic abundances and the cross sections
derived by \cite{Wilms_00} and \cite{Verner_96}, respectively.
To take into account the interstellar absorption we adopted the T{\"u}bingen-Boulder model ({\sc TBabs} in XSPEC). We fitted the spectrum using a Comptonized component  \cite[{\sc nthcomp} in XSPEC;][]{Zdi_96}. 
Finally, we added a constant to take into account the  different normalizations of the two instruments.
The initial model, called {\tt Model 1}, is defined as
$$
 \texttt{Model 1} = \textsc{Const*TBabs*nthcomp}.
 $$
We fixed at 0 the value of the parameter {\tt inp$\_$type} in {\sc nthcomp}, assuming    the seed-photon spectrum injected in the Comptonizing cloud to be a blackbody. 
The spectra A, B, C, and D were fitted simultaneously.
The value of the seed-photon temperature k$T_{bb}$ was tied 
at the same value in each spectrum and, finally, the value of the interstellar equivalent  neutral  hydrogen   column density  was tied at the same value in spectra C and D. 

\begin{figure*}
\centering
\includegraphics[scale=.6]{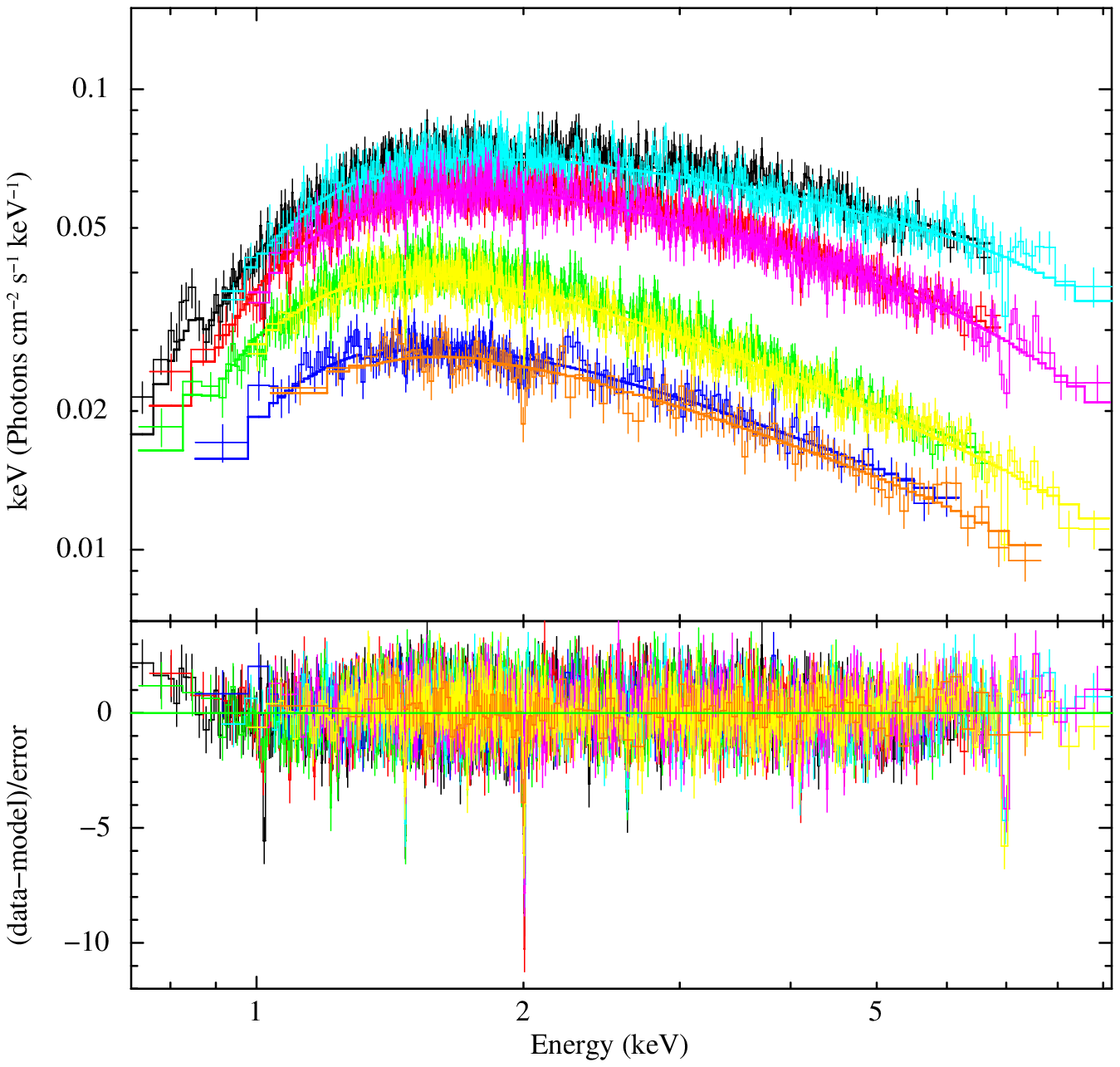}
\includegraphics[scale=.6]{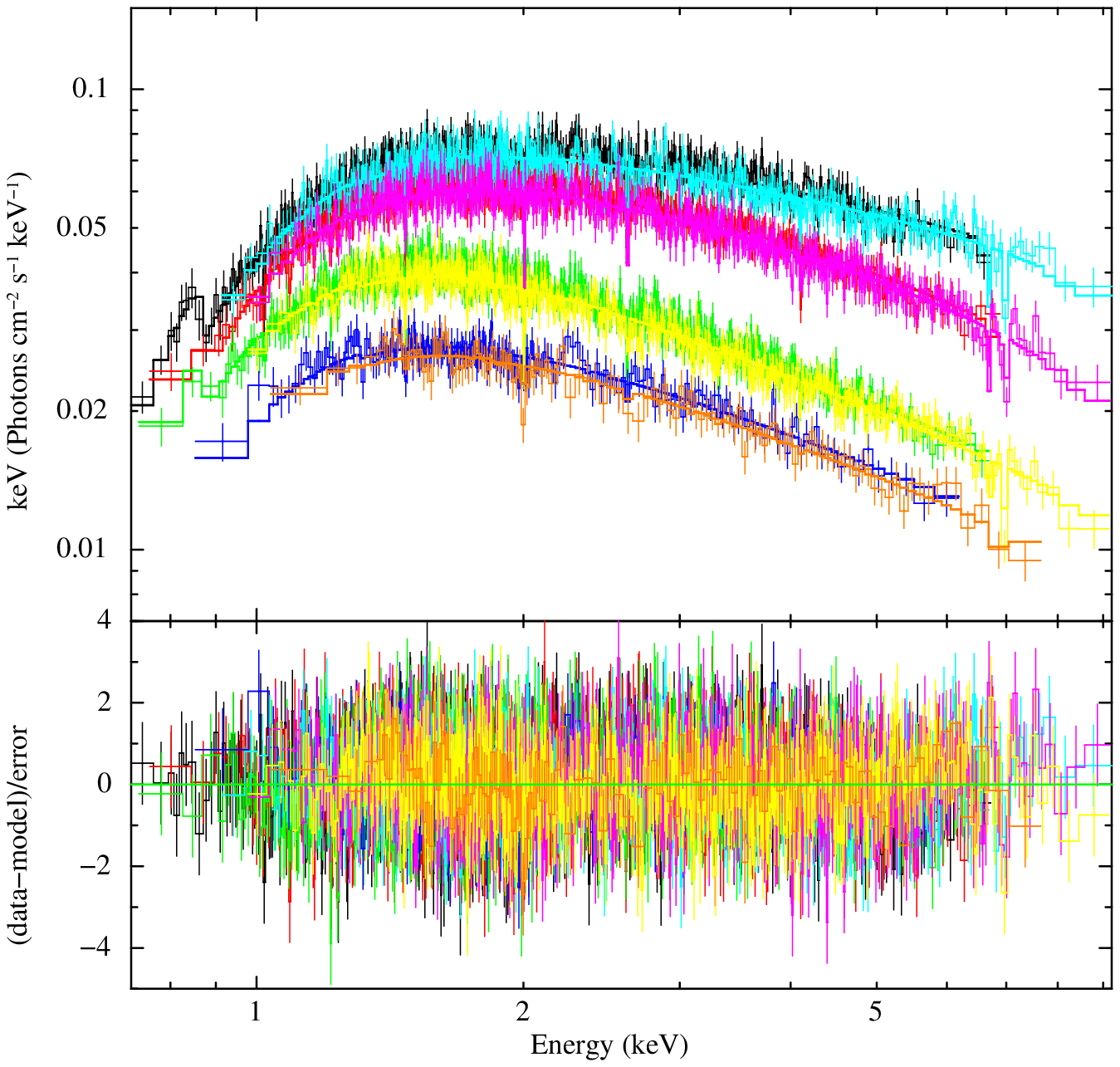}\\
\caption{Unfolded spectrum and residuals adopting {\tt Model 1} (left panels) and {\tt Model 2} (right panels). 
The black, red, green, and blue data correspond  to the 
first-order MEG of   spectra A, B, C, and D, respectively. The cyan, purple, yellow, and orange data   correspond to first-order  HEG spectrum of  spectra A, B, C, and D, respectively. The prominent absorption lines in residuals of {\tt Model 1} (bottom left panel) are associated with the presence of \ion{Ne}{x}, \ion{Mg}{xii}, \ion{Si}{xiv}, \ion{S }{xvi,} and \ion{Fe}{xxvi} ions.}
\label{figure:6}
\end{figure*}

We show the unfolded spectra and the residuals in the left panels  of Fig. \ref{figure:6}. The best-fit parameters
are shown in Table \ref{tab:mod1mod2}. Strong residuals compatible with  absorption lines are visible at   1 keV, 1.47 keV, 2 keV, 2.62 keV, and 6.97 keV.
Furthermore, below 0.9 keV there is a clear    excess in the residuals. 

\begin{table} 
\centering
\begin{threeparttable}
\caption{Identified Gaussian absorption lines.}
\scriptsize
\begin{tabular}{l@{\hspace{\tabcolsep}}c@{\hspace{\tabcolsep}}c@{\hspace{\tabcolsep}} c@{\hspace{\tabcolsep}}c@{\hspace{\tabcolsep}}}
\hline
\hline
Component &  Spectrum A & Spectrum B& Spectrum C&Spectrum D  \\

\hline\\
 & \multicolumn{4}{c}{\ion{Ne}{x} K$\alpha$ 1s-2p (1.0218 keV)} \\\\
 E (keV) & 
$1.020\pm 0.002$& 
$1.014\pm0.009 $ & 
$1.019^{+0.009}_{-0.019}$& 
-- \\
   $\sigma$ (eV) &     
  $<2.0$  & 
  $<17$ & 
  $<31$&
  -- \\
   Intensity{\textit{$\rm ^a$}}       &   
  $-8\pm2$& 
  $-8\pm5$& 
  $>-6$&
  --\\
  Eq. width{\textit{$\rm ^b$}} &     
  $-1.8\pm0.5$ & 
  $-2.1\pm0.8$ & 
  $-1.1\pm0.8$&
  --\\ 
  Bin size{\textit{$\rm ^c$}} (eV)&
  6 &
  10&
  7&
  --\\\\

& \multicolumn{4}{c}{\ion{Mg}{xii} K$\alpha$ 1s-2p (1.4723 keV)} \\\\
 E (keV) & 
$1.4715\pm0.0004$& 
$1.4720\pm0.0005$ & 
$1.4720^{+0.0005}_{-0.0009}$ &
--\\

  $\sigma$ (eV) &    
 $1.0\pm0.5$  & 
 $<1.8$ & 
 $<2.2$& 
 -- \\
 
Intensity     &   
$-5.9\pm0.7$&  
$-5.0\pm1.0$&
$-2.9\pm0.6$&
--\\

 Eq. width &   
 $-1.3\pm0.2$  & 
 $-1.1\pm0.2$ & 
 $-1.1\pm0.2$ &
 -- \\ 
 Bin size (eV)&
  2 &
  2&
  2&
  --\\\\

 & \multicolumn{4}{c}{\ion{Si}{xiv} K$\alpha$ 1s-2p (2.0055 keV)} \\\\
  E (keV) & 
 $2.0044\pm0.0004$ & 
 $2.0040^{+0.0007}_{-0.0002}$ & 
 $2.0039\pm0.0010$ & 
 $2.005^{+0.003}_{-0.014} $\\
 
  $\sigma$ (eV) &     
 $1.4^{+0.5}_{-1.2}$ & 
 $0.6^{+1.1}_{-0.6}$ & 
 $3.4^{+0.7}_{-1.2}$ & 
 $<29$\\
 Intensity     &   
 $-9.0\pm 0.8$& 
 $-6.7\pm0.8$& 
 $-5.2\pm0.8$& 
 $-2.2\pm1.5$\\
 Eq. width &      
 $-2.6\pm0.3$ & 
 $-2.2\pm0.2$ &
 $-2.8\pm0.3$ & 
 $-1.8\pm0.6$ \\ 
 Bin size (eV)&
  3 &
  3&
  3&
 20\\\\

 & \multicolumn{4}{c}{\ion{Si}{xiv} K$\alpha$ 1s-3p (2.376 keV)} \\\\
  E (keV) & 
 $2.372\pm0.002$ & 
 -- & 
 -- & 
 -- \\
 
  $\sigma$ (eV) &     
 $1.3^{+3.6}_{-1.3}$ & 
 -- &
 -- & 
 -- \\
 Intensity     &   
 $-6\pm 2$& 
 --& 
 --& 
 --\\
 Eq. width &      
 $-1.9\pm0.5$ & 
 -- &
 -- &
 -- \\
 Bin size (eV)&
 9 &
  --&
  --&
 --\\

 & \multicolumn{4}{c}{\ion{S}{xvi} K$\alpha$ 1s-2p (2.6217 keV) } \\\\
 E (keV) & 
$2.6157^{+0.0002}_{-0.0027}$& 
$2.621^{+0.002}_{-0.008}$ & 
$2.617^{+0.006}_{-0.004}$ &
-- \\

 $\sigma$ (eV) &      
$<3.0$ & 
$7\pm6$ &
$5\pm5$ 
&--\\

 Intensity      &  
$-9\pm 2$ & 
$-7\pm4$&
$-3.3\pm1.5$&
--\\ 

  Eq. width &     
 $-3.4\pm0.7$ & 
 $-3.3\pm0.9$ & 
 $-2.7\pm0.7$ & 
 --\\
 Bin size (eV)&
  8 &
  8&
  8&
 --\\\\

  & \multicolumn{4}{c}{\ion{Ca}{xx} K$\alpha$ 1s-2p (4.1050 keV)} \\\\
 E (keV) & 
$4.102\pm0.006$ & 
-- & 
-- &
--\\

  $\sigma$ (eV) &     
 $8^{+7}_{-8}$ & 
 -- & 
 --& 
 -- \\
 
 Intensity     &  
$-6\pm2$ & 
--&
--&
--\\

 Eq. width &    
 $-2.9 ^{+0.3}_{-2.5}$  & 
 -- & 
 -- &
 -- \\
 Bin size (eV)&
  20 &
  --&
  --&
 --\\\\

  & \multicolumn{4}{c}{\ion{Fe}{xxvi} K$\alpha$ 1s-2p (6.9662 keV)} \\\\
E (keV) & 
 $6.98^{+0.04}_{-0.09}$  & 
 $6.956^{+0.028}_{-0.003}$ &
 $6.96\pm0.04$ &
 -- \\
 
  $\sigma$ (eV) &    
 $<42$  &
 $<60$ &
 $<35$ &
 -- \\
 
 Intensity     &  
 $-16\pm6$ & 
 $-18\pm6$ &
 $-7\pm2$&
 --\\
 
  Eq. width &     
 $-13.1^{+1.0}_{-26.3}$  &
 $-33\pm7$ &
 $-43^{+23}_{-4}$ &
 -- \\
 Bin size (eV)&
 100 &
  100&
  120&
 --\\

\hline
\hline
\end{tabular}
      \begin{tablenotes}
\item[] The associated errors are at 68\% confidence level.      
 \item[a] The line intensity is in units of $\times 10^{-5}$ photons cm$^{-2}$ s$^{-1}$. 
 \item[b] The equivalent width is in units of eV. \item[c] The   size of the energy channel at the detected lines is estimated  from MEG spectra except for the \ion{Ca}{xx} and \ion{Fe}{xxvi} lines.
 \end{tablenotes}
\label{tab:line}
 \end{threeparttable}
\end{table}  
We fitted each observed absorption line with a  {\sc Gaussian} component. The energies, widths, intensities, and equivalent widths 
of each identified line are shown in Table \ref{tab:line} for each spectrum.  We detected the absorption lines 
associated with the presence of ions of \ion{Ne}{x},  \ion{Mg}{xii}, \ion{Si}{xiv}, \ion{S}{xvi}, \ion{Ca}{xx,} and \ion{Fe}{xxvi}. The line associated with the 
\ion{Ca}{xx} ions is detected only in spectrum A. Spectrum D only showed the line  associated with  \ion{Si}{xiv}, probably because of the low flux  of the spectrum.     For the sake of clarity the  sizes of the energy channels at which we identified the  absorption lines are shown in Table \ref{tab:line}. 

To   account for the presence of the prominent absorption lines,   to the    continuum  of  {\tt Model 1} 
we added the multiplicative
component {\sc zxipcf}\footnote{\scriptsize{https://heasarc.gsfc.nasa.gov/xanadu/xspec/models/zxipcf.html}}, which 
 takes into account a partial covering of ionized absorbing material. The component  reproduces the absorption from photoionized matter   illuminated by a power-law source with spectral index $\Gamma =2.2$, and it
assumes that the photoionized absorber has a microturbulent 
velocity of 200 km s$^{-1}$ \citep[see][for applications to AGN and Seyfert 1 galaxies]{Reeves_08,Miller_07}.
 Recently, this component was adopted to fit the absorbing features in the Fe-K region   associated with highly ionized matter surrounding the eclipsing NS-LMXB AX J1745.6-2901 \citep{Ponti_15}, the dipping source XB 1916-053 \citep{Gambino_19}, and the eclipsing source 
 MXB 1659-298 \citep{Iaria_19}. 
The parameters of this component are N$_{{\rm H}_{\xi}}$, log($\xi$), $f$, and $z$: N$_{{\rm H}_{\xi}}$   describes the equivalent hydrogen column density associated with the ionized absorber, log($\xi$)  describes the ionization degree of the absorbing material, $f$ is a  dimensionless covering fraction of the emitting region, and  $z$ is the redshift associated with  the absorption features. The parameter $\xi$  is defined as $\xi = L_x/(n_H r^2)$, where $L_x$ is the X-ray luminosity incident on the absorbing material, $r$ is the distance of the absorber from the X-ray source, and $n_H$ is the hydrogen atom density of the ionized absorber. 
 
To take into account that the ionized absorber can scatter the radiation out of the line of sight via Thomson or Compton scattering, we added the multiplicative component {\sc cabs} to the model. The only parameter of the component {\sc cabs} is $N_{H_{\rm cabs}}$, which describes the equivalent hydrogen column density associated with the scattering cloud. Finally, 
  to account for a partial covering of   the scattering material, we multiplied   {\sc cabs} by the component {\sc partcov}. This is a convolution model that allows us to convert  an absorption component into a partially covering absorption component, quantified by the  parameter  $f_{\rm cabs}$.  In order to make the model self-consistent, we tied  $f_{\rm cabs}$  to the parameter $f$  of the ionized absorber. Moreover, we linked   
  the equivalent hydrogen column density parameter  of the scattering cloud, $N_{H_{\rm cabs}}$, to the value  of the equivalent hydrogen column density of the ionized absorber, N$_{{\rm H}_{\xi}}$.  
While fitting simultaneously  spectra A, B, C and D,  we imposed   $f$, log($\xi$),  and the seed-photon temperature to assume the same value for each spectrum.    The model closely fits the absorption lines; however, the presence of an excess in the residuals below 0.9 persists. For this reason we added an absorption edge  with the energy threshold fixed at 0.871 keV to account for   the presence of \ion{O}{viii} ions along the line of sight, recently observed by \cite{Gambino_19}   analyzing the {\it Suzaku}    spectra of the source.

 The adopted model, hereafter called {\tt Model 2}, is defined as
\begin{equation*} 
\begin{split}
 \texttt{Model 2} = \textsc{Const*Edge*TBabs*(partcov*cabs)*}\\
 \textsc{{zxipcf*nthcomp}}. 
\end{split}
\end{equation*}

The addition of  {\sc zxipcf} and the absorption edge  improves the fits, with a  $\chi^2$(d.o.f.) of 3944(4542) and a $\Delta \chi^2$ of  1013 with respect to {\tt Model 1}.  We show 
the best-fit values in Table \ref{tab:mod1mod2} for each spectrum.  
The unfolded spectra and the corresponding residuals are shown in the right panels of Fig. \ref{figure:6};  the residuals associated with the absorption lines and the excess below 0.9 keV observed adopting {\tt Model 1} are now absent.

 We found  that the ionized absorber covers more than 86\% of the emitting source and the ionization parameter log($\xi$) is $4.33\pm0.02$.  The equivalent hydrogen column density associated with the ionized absorber 
N$_{{\rm H}_{\xi}}$ is $(19^{+7}_{-3})\times 10^{22}$, 
$(14^{+5}_{-2}) \times 10^{22}$, $(17^{+6}_{-3})\times 10^{22} $,
and $(10\pm7) \times 10^{22}$ cm$^{-2}$ for spectra A, B, C, and D, respectively. Furthermore, we observed a similar redshift value  for each spectrum (between $1.0\times10^{-3}$ and  $1.3\times10^{-3}$). 

The $\Gamma$ parameter of {\sc nthcomp} 
goes from  1.6 for spectrum A  to 1.8 for 
spectrum C and D, indicating that the spectral shape softens
 going from spectrum A to spectrum D. The electron temperature of the Comptonizing cloud is compatible with  3-4 keV for all the spectra even if it is not well constrained for spectra C and D,  while the seed-photon temperature is
$0.16 \pm 0.04$ keV.  

Finally, the best-fit value of the optical depth of the absorption edge is $0.14\pm0.05$ and the equivalent hydrogen column density of the interstellar matter is compatible with $ 0.5 \times 10^{22}$ cm$^{-2}$ for each spectrum.

Assuming a distance to the source of 8.9 kpc \citep{Galloway_08},  the unabsorbed luminosity in the 0.1-100 keV  energy range is $1.46 \times 10^{37}$ erg s$^{-1}$,  $1.0 \times 10^{37}$ erg s$^{-1}$, $0.67 \times 10^{37}$ erg s$^{-1}$, and 
$0.41  \times 10^{37}$ erg s$^{-1}$ for A, B, C, and D, respectively.

\section{Spectral analysis of the dips}
To analyze the average spectrum during the dips we 
 extracted the events from  observations 20171, 20172, 21106, 21662, 21663, 21664, and  21666.   The dip events were selected from observation 20171 taking into account
the good times intervals (GTIs) obtained from the first-order MEG+HEG light curve in the 0.3-10 keV energy range where the count rate was lower than 3 c s$^{-1}$. Similarly, we used the threshold of 2 c s$^{-1}$ for ObsID. 20172, 2.4 c s$^{-1}$  for ObsID. 21662, 2 c s$^{-1}$ for ObsID. 21663 and 21664,  1.5 c  s$^{-1}$  for ObsID. 21106, and  1.1 c s$^{-1}$  for ObsID. 21666. We extracted 
the first-order HEG and MEG spectrum from each observation and
combined them  using the CIAO script {\tt combine$\_$grating$\_$spectra}. 
Each spectrum was then grouped to have   at least  100   counts per energy channel; the resulting spectrum has an exposure time of 
9.5 ks. The energy range adopted for the spectral analysis is 0.9-7 keV and 1-9 keV  for MEG and HEG, respectively. 
 
We fitted the spectrum adopting {\tt Model 2}; 
however, we excluded the {\sc edge} component because of the  low flux of the spectrum. does not allow us to constrain it. We kept     the photon-index value,
the electron temperature, and the seed-photon temperature of the  {\sc nthcomp} component fixed  to 1.8, 4.3 keV, and 0.16 keV, respectively. We did this  so that the dip-spectrum was extracted from the observations belonging to  set C and D, under  the assumption that the spectral shape of the continuum emission is similar both inside and outside the dips.  

\begin{figure}
\includegraphics[scale=.64]{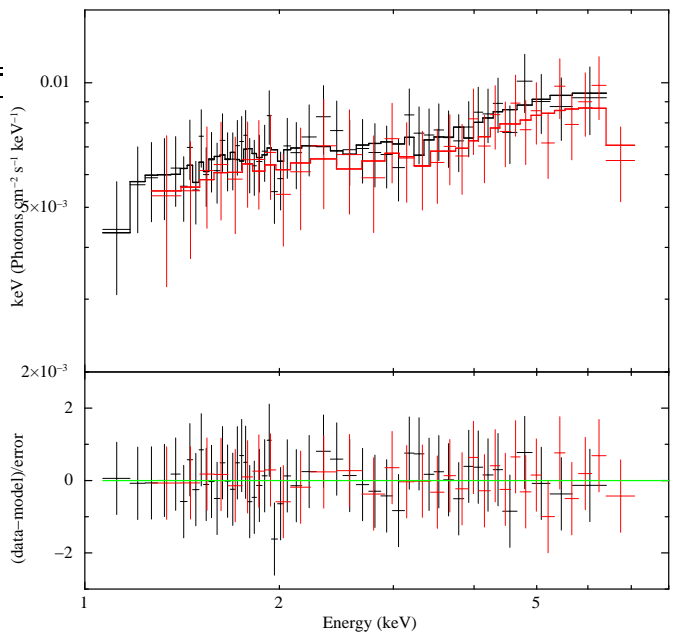}
\caption{Unfolded spectrum and residuals of the dip spectrum
adopting {\tt Model 2}. The black and red data correspond to the first-order MEG and HEG spectra, respectively.}
\label{figure:7}
\end{figure}

We show the  best-fit parameters in Table \ref{tab:dip}; the unfolded spectrum and the corresponding residuals are shown in Fig. \ref{figure:7}.
\begin{table} 
\centering
\begin{threeparttable}
\caption{Best-fit parameters of the spectrum during the dip.}
\begin{tabular}{l@{\hspace{2pt}}l@{\hspace{\tabcolsep}}l@{\hspace{\tabcolsep}}}
\hline
\hline

  &   & Dip spectrum\\
 \hline 
Model & Component &   \\      
      
\hline

{\sc Constant} &C$_{\rm HEG}$  &  
$0.92\pm0.08$ \\\\

{\sc partcov} & $f_{\rm cabs}${\textit{$\rm ^a$}} & $>0.68$  \\
      {\sc cabs} & N$_{\rm H_{\rm cabs}}${\textit{$\rm ^b$}} ($\times 10^{22}$)  &     
      $61^{+56}_{-10}$  \\\\         
 {\sc zxipcf} & N${_{\rm H}}_{\xi}$ ($\times 10^{22}$)   &  $61^{+56}_{-10}$        \\

&log($\xi$)& $2.8^{+0.2}_{-0.4} $\\
& $f$ & $>0.68$      \\  
& Redshift& 0 (fixed) \\ \\

{\sc TBabs} & N$_{\rm H}$($\times 10^{22}$)    &   $0.9\pm0.2$ \\\\             

{\sc nthcomp}     &    $\Gamma$ &  
1.8 (fixed)   \\

     & kT$_{bb}$ (keV)  &    0.16 (fixed) \\

     & kT$_{e}$ (keV)  &   4.3 (fixed)    \\

 & Norm ($\times 10^{-2}$) &      
 $7.2\pm1.4$ \\\\

 & $\chi^2/dof$ &    
 18.3/76 \\ 
\hline
\hline
\end{tabular}
      \begin{tablenotes}
\item[a] The value of the parameter is tied to the value of $f$. 
\item[b] The value of the parameter is tied to the value of  N$_{{\rm H}_{\xi}}$. 
 \end{tablenotes}
\label{tab:dip}
 \end{threeparttable}
\end{table}  
We found that during the dip   the  ionized absorber is characterized by log($\xi$)$= 2.8^{+0.2}_{-0.4}$, suggesting  a low   ionization of the absorber. The equivalent  hydrogen column density 
 N${_{\rm H}}_{\xi}$ associated with the ionized absorber is  $(61^{+56}_{-10})\times 10^{22}$ cm$^{-2}$, which is larger than  the values obtained  during the persistent emission by a factor between 3 and 6.

The covering fraction is $>0.68$, compatible with the values  obtained for spectra A, B, C, and D.
 The equivalent hydrogen column density  associated with the interstellar matter is $(0.9\pm0.2)\times 10^{22}$ cm$^{-2}$. This value is larger than the observed value during the persistent emission. We cannot exclude that a partial contribution to this value comes from local neutral hydrogen in the hypothesis that during the dip the absorber is composed of neutral and mildly ionized matter.   
 
Finally,  the 0.1-100 keV unabsorbed luminosity, assuming a distance to the source of 9 kpc, is  $0.57 \times 10^{37}$ erg s$^{-1}$.

\section{Discussion}
From the analysis of the {\it Chandra}  and  {\it Swift/XRT} observations we inferred three new dip arrival times that,  added to the previous 27 reported  by 
\cite{iaria_15}, allowed us to update the orbital ephemeris of XB 1916-053 by extending the temporal baseline from 36 to 40 years. The largest baseline   excludes some solutions previously 
plausible with the available data, such as the LQ ephemeris shown by  \cite{iaria_15}. 
We find that the LQS ephemeris  better models the dip arrival times,  
improving the constraints on 
  the orbital period derivative and  the  sinusoidal modulation. 
 The orbital period derivative is $\dot{P}=1.46(3)\times 10^{-11}$ s s$^{-1}$ and the sinusoidal modulation has a period of $P_{\rm {mod}}=9099\pm302$ d.  
 
\cite{iaria_15} discussed that such a large orbital period derivative  
 can be explained only assuming that the mass transfer rate
from the CS to the NS is highly non-conservative. The authors estimated that more that 90\%  of the mass transfer rate from the CS is lost by the system, the mass ratio $q=m_2/m_1$  
 ($m_1$ and $m_2$ are the NS and CS mass in units of solar masses) is close to 0.013 and   the NS mass should be larger than 2.1 M$_{\odot}$.

Here we show  that the conclusions of \cite{iaria_15} were  strongly constrained by the hypothesis  that the matter 
leaves the system from the inner Lagrangian point. To discuss this point  we have to estimate   the mass ratio $q$ of XB 1916-053. 

\subsection{Mass ratio $q$ of XB 1916-053}
XB 1916-053 shows a superhump period $P_{sh}$ \citep{Chou_01} and a negative superhump period, also called infrahump period,  $P_{ih}$ \citep{Retter_02, Hu_08}.  
\cite{Chou_01} discussed that the optical modulation close to 3028 s is likely caused by the coupling of the orbital motion with a 3.9-day disk apsidal precession period, like the superhumps in    SU UMa-type dwarf novae,  obtaining from this assumption $q \simeq 0.022$, while \cite{Hu_08} discussed the infrahump period of 2979.3(1.1) s detected by \cite{Retter_02} as 
being due to the  nodal precession period  of the tilted accretion disk, and finding  $q \simeq 0.045$. 
Considering that the orbital period of the system is close to 3000.66 s, the apsidal precession period of the disk should be  3.9087(8) d, while the retrograde nodal precession period   4.86 d. However, the reported values of $q$  do not depict a self-consistent scenario.   
We show below that the value of $q=0.048$ is consistent with the detected values of  $P_{sh}$  and $P_{ih}$ only assuming an outer radius of the disk   truncated at a 3:1 resonance.

 \cite{Hirose_90},  performing hydrodynamic simulations of accretion disks to study the superhump phenomenon in SU UMa stars,   showed that the  apsidal precession  frequency $\omega_p$ can be written as $\omega_p/\omega_{\rm orb}= Z(r) q/(1+q)^{1/2}$, where 
\begin{equation*} 
\begin{split}
Z(r)=\frac{1}{2}\frac{1}{r^{1/2}}\frac{d}{dr}\left[ r^2\frac{d}{dr}B_0(r)\right],
\end{split}
\label{eq:a}
\end{equation*} 
and 
\begin{equation*} 
\begin{split}
B_0(r)=\frac{1}{2}b_{1/2}^{(0)}=1+\frac{1}{4}r^2+\frac{9}{64}r^4+...,
\end{split}
\label{eq:laplace}
\end{equation*} 
which is the Laplace coefficient of order 0 in celestial mechanics
\citep[see ][chapter 15, eq. 42]{Brouwer_61} and $r$ is the ratio of the   accretion disk radius $r_{\rm disk}$ to the orbital separation $a$.  Combining the   expressions 
we find that
\begin{equation} 
\begin{split}
\frac{\omega_p}{\omega_{\rm orb}}  =\frac{q}{(1+q)^{1/2}}r^{3/2}\left[ 0.75+\frac{45}{32}r^2\right].
\end{split}
\label{eq:osaki}
\end{equation}

\cite{Hirose_90}  demonstrated that the tidal instability in the  accretion disks is caused by the resonance between the  particle orbits in the disk and the companion star with a 3:1 period ratio in SU UMa dwarf novae. Since   $r=r_{\rm disk}/a$ can be expressed by the  equation
\begin{equation} 
\begin{split}
\frac{r_{\rm disk}}{a} =\frac{R_{jk}}{a} =\left(\frac{j-k}{j}\right)^{2/3} (1+q)^{-1/3},
\end{split}
\label{eq:rdisk}
\end{equation} 
with $j=3$ and $k=2$ for a 3:1 resonance \citep[see][eq. 5.125]{Frank_02}, we can combine 
 the two last equations to obtain  
\begin{equation} 
\begin{split}
\frac{\omega_p}{\omega_{\rm orb}} =  \frac{1}{3}\frac{q}{1+q}[0.75+0.325(1+q)^{-2/3}];
\end{split}
\label{eq:ompomorb}
\end{equation} 
using the value $\omega_p/\omega_{\rm orb}\simeq 0.0089$
inferred by \cite{Chou_01} we find  $q\simeq 0.0256$. 

However, \cite{Hirose_90}  estimated the apsidal precession period of the disk  taking into account only the dynamical effects of the matter.
\cite{Lubow_91a,Lubow_91b,Lubow_92} showed that the apsidal precession rate $\omega$ for an eccentric disk is given 
by three terms    $\omega=\omega_{\rm dyn}+\omega_{\rm press}+\omega_{\rm tr}$, 
where  $\omega_{\rm dyn}$ is the dynamical precession frequency discussed by \cite{Hirose_90}  and shown above, $\omega_{\rm press}$ is a term due to the pressure, and 
 $\omega_{\rm tr}$ is a transient term related to the time-derivative of the mode giving rise to the dynamical precession that 
 can be neglected in steady state.  Since XB 1916-053 is a persistent source that does not show outburst, and since \cite{Callanan_95}
 verified the stability of the optical period over seven years,  we do not consider the last term as we assume 
 that the disk  precesses in a steady state.  
  
  \cite{Lubow_92} showed that the pressure term can be written as $\omega_{\rm press}$=$-k^2c_s^2/(2\Omega)$, where 
  $k$ is the radial wavenumber of the mode, $c_s$ the sound speed of the gas, and $\Omega$ is the frequency of a particle in the disk at a given radius. The negative sign
  is due to the pressure
term associated with a spiral arm in the disk, which acts in the opposite sense with respect to the dynamical term.
 For a spiral wave, the pitch angle $\theta$ is related to $k$ by the relation $\tan \theta = (kR)^{-1}$, where $R$ is the distance from the central source. 

 \cite{Pearson_06}, considering a 3:1 resonance and adopting   eq. \ref{eq:rdisk}, expressed the term associated with the pressure of the spiral wave  $\omega_{\rm press}$ as
  \begin{equation} 
\begin{split}
\frac{\omega_{\rm press}}{\omega_{\rm orb}} =   -\frac{j^{1/3}}{2}(1+q)^{2/3}\left( \frac{c_s}{\omega_{\rm orb}\; a} \tan^{-1}\theta \right)^2 = - j^{1/3} \eta_A (1+q)^{2/3}, 
\end{split}
\label{eq:pearson}
\end{equation} 
where $a$ is the orbital separation, $j=3$,  and $\eta_A = 0.5 [c_s/(\omega_{\rm orb}\; a)]^2 \tan^{-2}\theta$. We combined   eqs. \ref{eq:ompomorb} and  \ref{eq:pearson} and obtained 
 \begin{equation} 
\begin{split}
\frac{\omega}{\omega_{\rm orb}}=\frac{1}{3}\frac{q}{1+q}[0.75+0.325(1+q)^{-2/3}]- 3^{1/3} \eta_A (1+q)^{2/3}, 
 \end{split}
\label{eq:mia}
\end{equation}  
where $\omega$ is now the apsidal precession frequency of the disk taking into account  the pressure term.

\cite{Hu_08}, assuming a 3:1 resonance, inferred a value of $q \simeq 0.045$ based on the negative superhump with a period of 2979.3 s observed by 
\cite{Retter_02}, and interpreted as the beat period between the orbital period and the 4.86 d  nodal precession period of the disk. The value of $q \simeq 0.045$  was obtained from the 
expression proposed by \cite{Larwood_96} and \cite{Montgomery_09}
\begin{equation} 
\begin{split}
\frac{\omega_n}{\omega_{\rm k}} =   -\frac{15}{32}qr^3\cos\delta,
\end{split}
\label{eq:larwood}
\end{equation} 
where $\omega_n$ is the angular frequency associated with 
the nodal precession of the disk,  $\omega_k$ is the Keplerian frequency of the matter at a given radius $r$  of the accretion disk  
that is defined in units of orbital separation $a$, and   the angle $\delta$  describes the orbit of the CS with respect to the NS. For $\delta \simeq 0$ the orbits of the two bodies are coplanar   \citep{papa_95}. 
 The minus sign is present because the nodal frequency is retrograde with respect to the orbital frequency \citep[see Fig. 6.18 in][]{Hellier_01}.

Rewriting the Keplerian frequency in the  known form
$\omega_k=(GM_1/R^{3})^{1/2}$ and combining it with Kepler's third law $a^3=GM_1(1+q)/\omega_{orb}^2$ we find 
$\omega_k=\omega_{orb}(1+q)^{-1/2}r^{-3/2}$. We combined the last expression with eq. \ref{eq:larwood}, obtaining
\begin{equation} 
\begin{split}
\frac{\omega_n}{\omega_{\rm orb}} =   -\frac{15}{32}q(1+q)^{-1/2}r^{3/2}\cos\delta.
\end{split}
\label{eq:larwood2}
\end{equation} 
Assuming that $r=r_{disk}/a$ \citep[the same assumption made by][]{Hu_08} and combining the last equation with eq. \ref{eq:rdisk} for a 3:1 resonance, we infer
\begin{equation} 
\begin{split}
\frac{\omega_n}{\omega_{\rm orb}} =   -\frac{5}{32}\frac{q}{1+q}\cos\delta.
\end{split}
\label{eq:larwood3}
\end{equation} 
Because $\omega_n/\omega_{orb} \simeq 7.146 \times 10^{-3}$, we adopt the nodal period of 4.86 days and  find that $q\simeq 0.048$ for $\cos\delta\simeq1$.

\begin{figure}
\includegraphics[scale=.50]{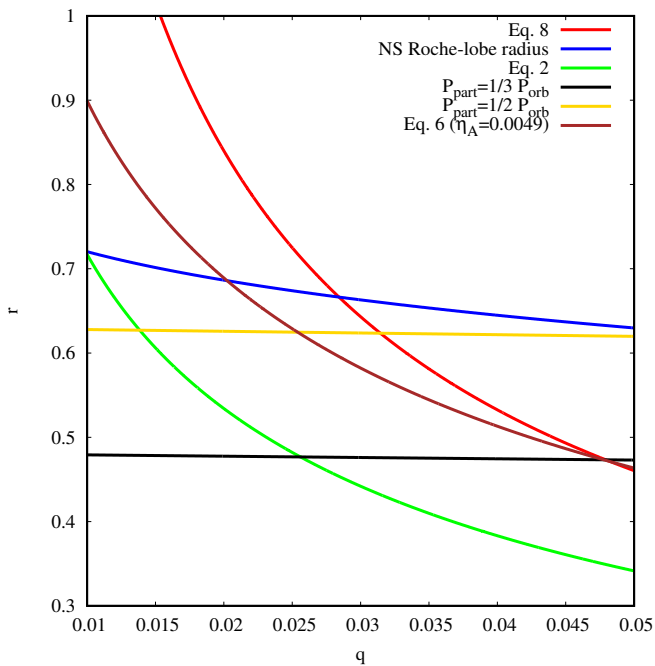}
\caption{Characteristic radii of XB 1916-053 in units of orbital separation vs the mass ratio q. Show are the NS Roche-lobe radius (blue curve), the  radius of the disk at which   a 2:1 and 3:1 resonance could occur (yellow and black curves), the radius associated with the apsidal rate of the disk for a 3:1 resonance without and with the pressure term due to the presence of a spiral wave in the disk (green and brown curve), and the radius associated with  the nodal precession period for a 3:1 resonance (red curve).}
\label{figure:inconsisyency}
\end{figure}

We show in Fig. \ref{figure:inconsisyency} 
the Roche-lobe radius of the NS as a function of $q$  (blue curve), 
the outer radius of the accretion disk for a 2:1 resonance (yellow curve) and for a 3:1 resonance (black curve). 
The green curve represents the outer radius of the disk 
adopting the value of  $\omega/\omega_{\rm orb} \simeq 8.9 \times 10^{-3}$ 
obtained by \cite{Chou_01}, who neglect  the pressure term due to the  spiral wave in the apsidal precession rate (see eq. \ref{eq:ompomorb}). 
The red curve represents the outer radius of the disk assuming a nodal precession of 4.86 days  obtained from   eq. \ref{eq:larwood3}.  All the radii are in units of orbital separation $a$. 

Initially, we note that the green and red curve do not have   intersections that exclude any self-consistent scenario; in other words,  the outer radius of the disk estimated from the observed apsidal precession period is not compatible with that obtained from the observed nodal precession period  if we ignore the pressure term.  Taking into account $\omega_{\rm press}$, the outer radius of  
the disk obtained from the apsidal precession period can intersect the red curve. It is possible to obtain a  pair of $r$ and $q$  values for a 3:1 resonance assuming that the parameter $\eta_A\simeq 0.0049$. For that value we find 
$q \simeq 0.048$ and an outer disk radius   $r\simeq 0.474$ in units of orbital separation. It could be possible 
a solution for a 2:1 resonance for which $q \simeq 0.0315$, $r\simeq 0.62$, and $\eta_A\simeq 0.0089$; however, since the truncation radius of the disk due to the tidal interaction with the CS is $r_T=R_T/a=0.6/(1+q)$ \citep[for $q$ between 0.03 and 1;][]{Pac_77}, we obtain $r>r_T$, making the solution unrealistic.

A further indication  that the mass ratio $q$ could be close to 0.048 is  given by   the empirical expressions inferred for the   CVs, in which the period excess $\epsilon$ depends on $q$. 
Using the values of the superhump period  $P_{\rm sh}=3027.5510(52)$  s and the orbital period $P_{\rm orb}=3000.6508(9)$ \citep[see Table 1 in][and references  therein]{Retter_02} we can define the period excess as 
 \begin{equation} 
\begin{split}
\epsilon =  \frac{P_{\rm sh}-P_{\rm orb} }{P_{\rm orb}}=(8.965\pm0.002) \times 10^{-3}.
 \end{split}
\label{eq:mia}
\end{equation}  
Using the empirical form $\epsilon = 0.18q+0.29^2$ \citep{Patterson_05} we obtain  $q\simeq 0.045$.  \cite{Goodchild_06} inferred $\epsilon =0.2076(3)q -4.1(6) \times 10^{-4} $, where  the  small offset from the origin is due to pressure-induced retrograde precession of a stable eccentric mode in systems with very low $q$.  Using the last relation we infer that $q = 0.0452 \pm 0.0004$. 

\begin{figure}
\includegraphics[scale=.50]{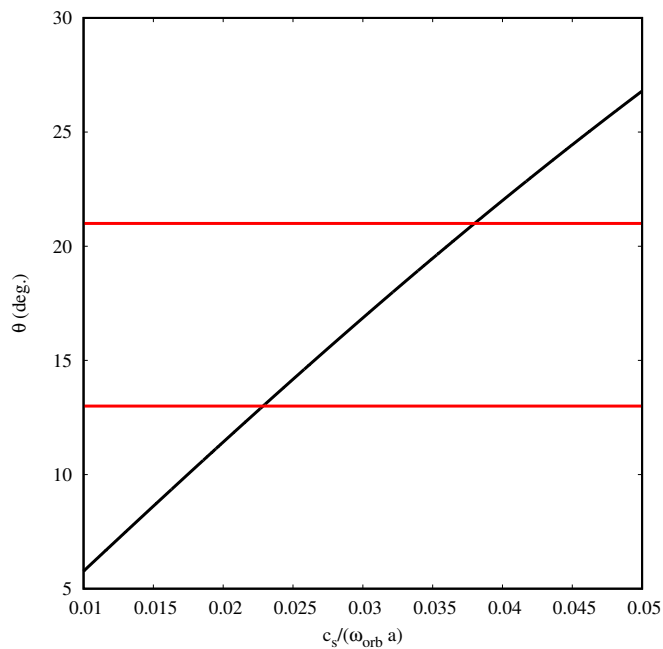}
\caption{Pitch angle of the spiral wave vs the normalized sound of speed  (black curve). The red horizontal lines correspond to the angles of pitch between 13$^{\circ}$ and 21$^{\circ}$, as obtained by \cite{Montgomery_01} from simulations.}
\label{figure:eta}
\end{figure}

We obtain  $q \simeq 0.048$ for 
\begin{equation} 
\begin{split}
\eta_A=   \frac{1}{2} \left(\frac{c_s}{\omega_{\rm orb}\; a}\right)^2  \tan^{-2}\theta=0.0049.
 \end{split}
\label{eq:eta}
\end{equation} 
 \cite{Lubow_92}  gave a range of 0.01-0.05 for the normalized speed of sound, 
$ c_s/(\omega_{\rm orb}\; a),$ and inferred that for a tightly wrapped spiral arm 
the pitch angle $\theta$ is between 5.7$^{\circ}$ and 31$^{\circ}$. We show in Fig. \ref{figure:eta} that for $\eta_A\simeq0.0049$ this condition
is verified, and we obtain that the pitch angle is between
5.8$^{\circ}$ and 26.7$^{\circ}$.  From simulations \cite{Montgomery_01} restricted the range of the pitch angle 
between 13$^{\circ}$ and 21$^{\circ}$; these values  restrict  the range 
of the normalized sound of speed between 0.023 and 0.038. 
Using  Kepler's third law and assuming $q=0.048$  we obtain that the sound speed $c_s$ is between 
$15.2 m_1^{1/3}$ km s$^{-1}$ and $25.2 m_1^{1/3}$ km s$^{-1}$.   This range of values is compatible 
with a disk temperature   in the outer region  between $3\times10^4$ K and  $7\times10^4$ K \citep[see eq. 2.21 in][]{Frank_02}, which  is compatible with the results obtained by    \cite{Nelemans_06} from the analysis of the optical band of XB 1916-053. The authors found that 
a LTE model consisting of pure helium plus overabundant nitrogen closely fits the observed spectrum finding that the  model has a temperature of  $\sim 3\times10^4$ K.

\subsection{Neutron star mass $m_1$}

Assuming that the CS fills its Roche lobe, we can obtain an indication of the NS mass. The  CS is a degenerate star and its radius $R_2$ 
is given by $R_2/R_{\odot} = 0.0126 (1+X)^{5/3} m_2^{-1/3}$, 
where $X$ is the fraction of hydrogen in the star. This  equation has to be corrected for the thermal bloating factor $b_f\ge1$, which is the ratio of the CS radius to the radius of a star with the same mass and composition; this star is completely degenerate and supported only by the Fermi pressure of the electrons.
Adopting 
the Roche-lobe prescription of \cite{Eggleton_83},   
 using Kepler's third law and  imposing that   $R_{\rm l2}= b_f R_2$,  where $R_{\rm l2}$ is the Roche-lobe radius of the CS, we obtain
\begin{equation} 
\begin{split}
m_1
= 
14.17 \; q^{-1/2}(1+q)^{-1/2} \left[\frac{0.49\;q^{2/3}}{0.6 q^{2/3}+\ln(1+q^{1/3})} \right]^{-3/2}\\P^{-1} (1+X)^{5/2} b_f^{3/2},
\end{split}
\label{eq:m1}
\end{equation} 
where $P$ is the orbital period in seconds. 

\cite{Heinke_13} inferred   $X \sim 0.14$ for a NS mass of 1.4 M$_{\odot}$, compatible with the measurements of \cite{Nelemans_06}. 
We can predict how $X$ varies with respect to $m_1$.  
Even though the CS is almost a pure helium dwarf, the small percentage of hydrogen  that can be  transferred onto the NS via Roche-lobe accretion can  significantly change the energy released due to hydrogen’s larger energy release per nucleon \citep[energy released per nucleon $Q_{\rm nuc} = 1.6+4.0 X$ MeV nucleon$^{-1}$;][]{Cumming_03}. \cite{Galloway_08}, using {\it RXTE} data, measured the ratio $\alpha$ of burst to persistent flux 
between two consecutive type I X-ray bursts temporally spaced by 6.3 hours obtaining $\alpha=78.8\pm0.3$. We rewrite   eq. 6 shown 
by \cite{Galloway_08} as
\begin{equation} 
\begin{split}
\alpha
= 34.57\; m_1 (0.4+X)^{-1},
\end{split}
\label{eq:burst}
\end{equation} 
in which we assumed a NS radius of 10 km. Adopting $\alpha=78.8\pm0.3$ we find that $X$ is 0.21, 0.3, and 0.39   for a NS mass of 1.4, 1.6, and 1.8  
M$_{\odot}$ respectively, while  for a NS mass of 2.1 M$_{\odot}$ we find that 
$X \simeq 0.52$, in contrast with    the   pure helium dwarf nature of the CS. Furthermore, assuming an orbital period value $P=3000.66$ s and a mass ratio $q=0.048$, and combining 
  Eqs. \ref{eq:m1} and \ref{eq:burst}, we find 
\begin{equation} 
\begin{split}
b_f
= m_1^{2/3} (0.376+0.275 m_1)^{-5/3}.
\end{split}
\label{eq:bf}
\end{equation} 
We obtain that $b_f$ is 1.97, 1.92, and 1.86 for a NS mass of 1.4, 1.6, and 1.8  M$_{\odot}$, respectively.
In the following we explore the NS mass between 1.4 and 1.8 M$_{\odot}$ in order to have a fraction of hydrogen $X$ in the CS lower than 0.4. 

Finally, we can express $m_2$ from  eq. \ref{eq:m1}   remembering 
 that $m_1=m_2q^{-1}$. We find
 \begin{equation} 
\begin{split}
m_2
= 
41.31 \left(\frac{q}{1+q}\right)^{1/2} [0.6+q^{-2/3}\ln(1+q^{1/3})]^{3/2}\\P^{-1} (1+X)^{5/2} b_f^{3/2}.
\end{split}
\label{eq:m2}
\end{equation} 
\begin{figure*}
\centering
\includegraphics[scale=.34]{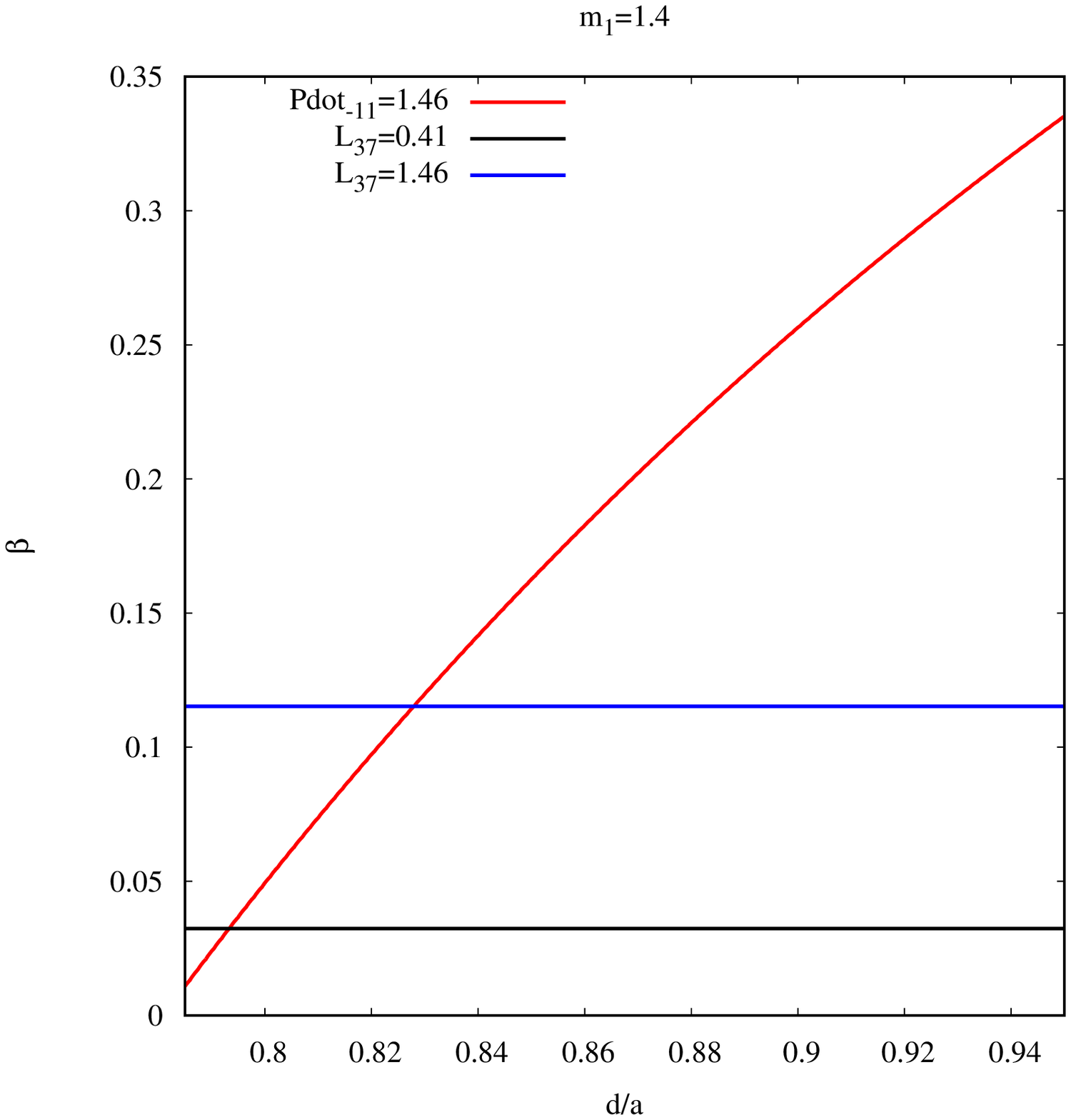}
\includegraphics[scale=.34]{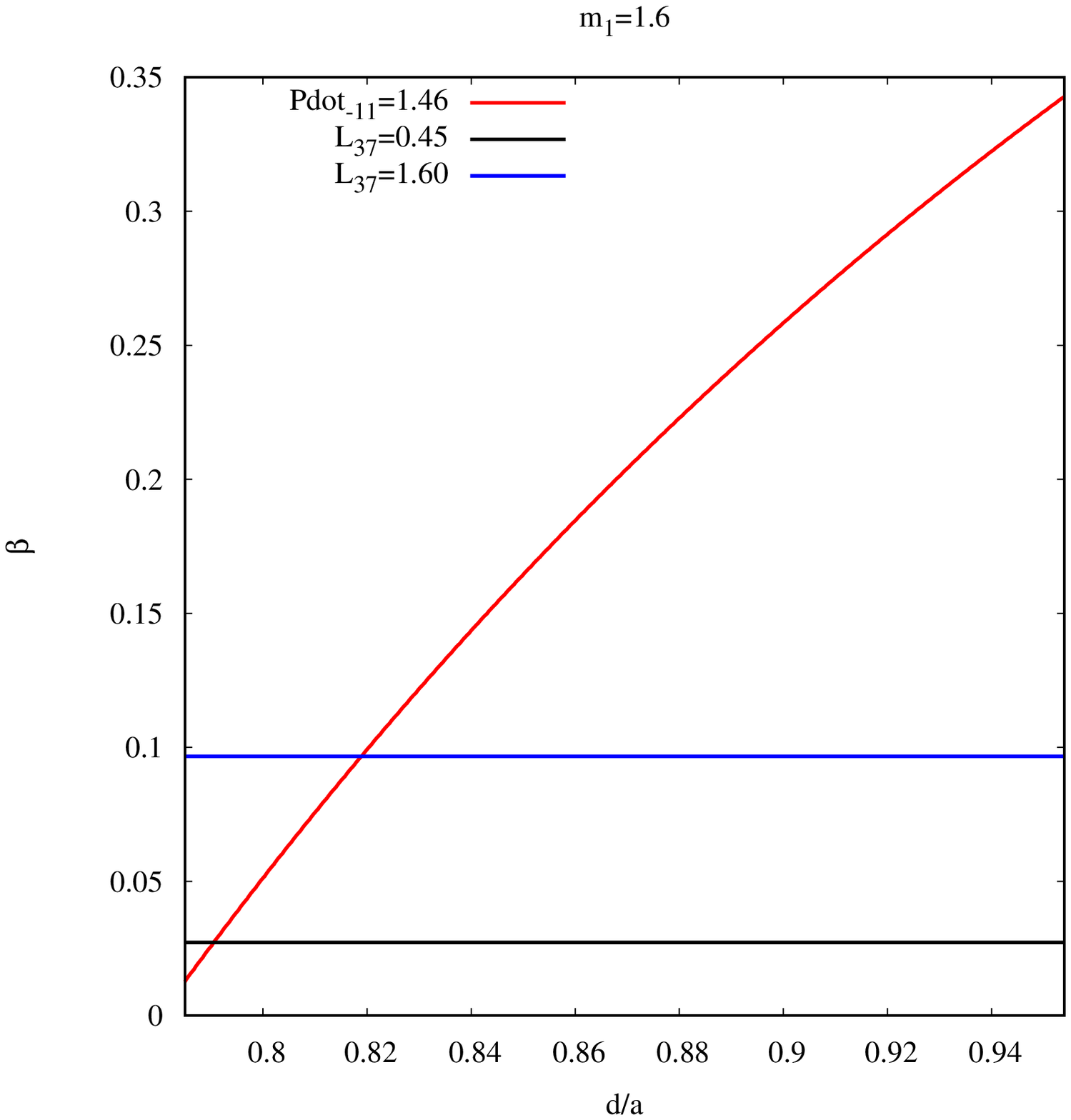}
\includegraphics[scale=.34]{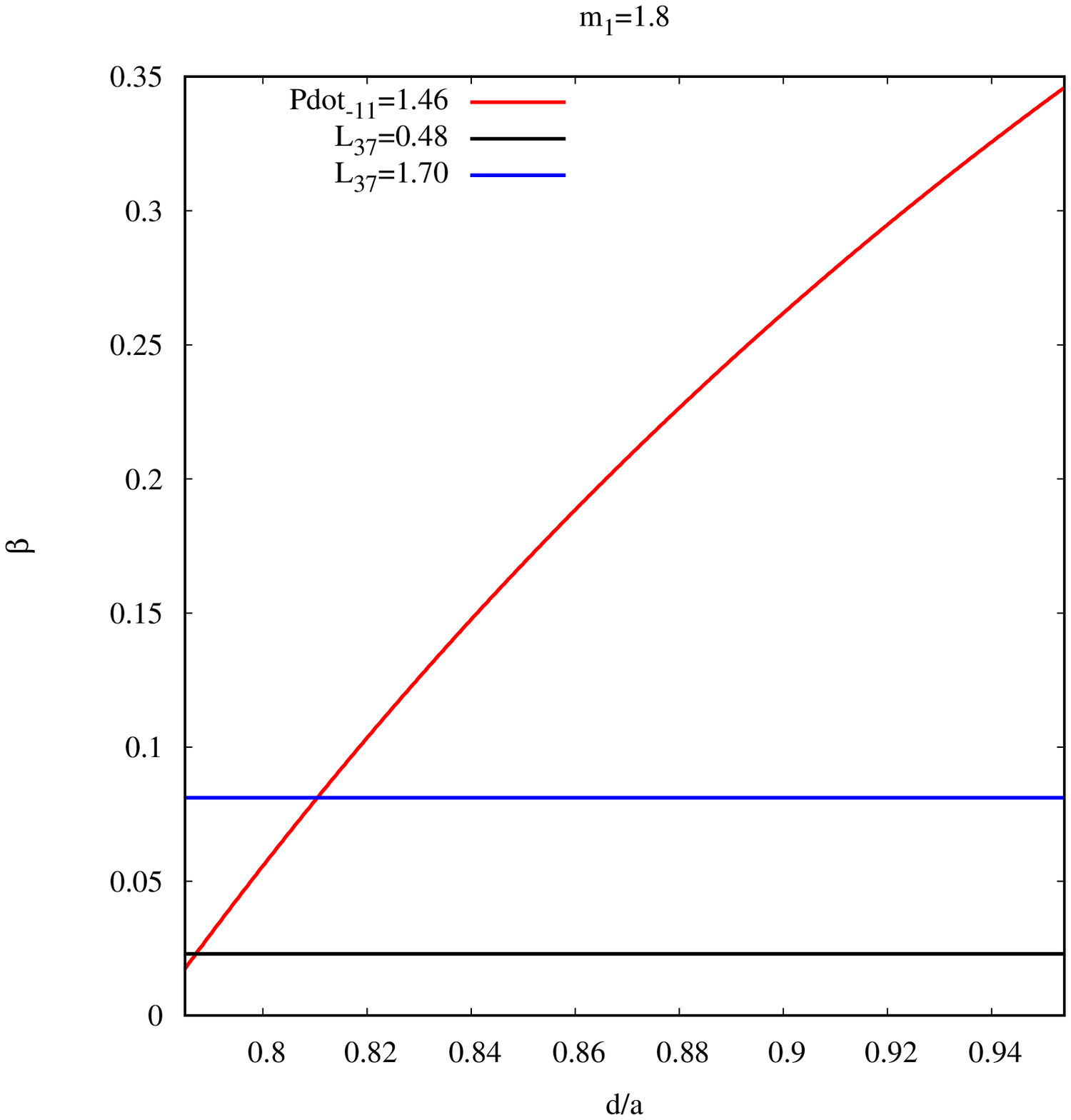}\\
\caption{Constraint of the matter-ejection point position and $\beta$ for $q=0.048$. The left, middle, and right panels describe the constraints for $m_1$ values 1.4, 1.6, and 1.8 M$_{\odot}$. The red curve 
represents the orbital period derivative of 
$1.46 \times 10^{-11}$ s s$^{-1}$ for the pairs $\beta$-$d/a$. The black and blue horizontal lines represent the minimum and maximum luminosity of the source during the observations. The minimum and maximum value of $d/a$ correspond to the inner Lagrangian point and the CS position with respect to the center of mass. The observed luminosities suggest that the mass transfer is highly non-conservative and that the ejected matter leaves the system close to the inner Lagrangian point.}
\label{figure:9}
\end{figure*} 

\subsection{Orbital period derivative and   luminosity of   XB 1916-053}

Since XB 1916-053 has an orbital period close to 50 min we assume that the angular momentum lost via magnetic braking 
is negligible with respect to the angular momentum lost via gravitational waves. Combining  eqs. 1, 2, and 3 shown by  
\cite{Rappaport_87} with the assumption that the stellar adiabatic   index $\xi_{\rm ad}$ is -1/3, as suggested by the authors, we obtain
\begin{equation} 
\begin{split}
-\frac{\dot{m_2}}{m_2} = \frac{6.046\times 10^{-16}}{\Lambda(q,\alpha,\beta)} m_1^{5/3} q (1+q)^{-1/3}P_h^{-8/3} {\rm \; s^{-1}},
\end{split}
\label{eq:m2dotonm2_1}
\end{equation} 
where $P_h$ is the orbital period in   hours, 
$\beta$ is the fraction of the mass that transferred  
from the CS accretes onto the NS and $\alpha$ is linked to 
the rate of specific angular momentum $\dot{J}_{\rm mass}$ lost via mass loss  by the binary system. Defining 
$\dot{J}_{\rm mass}=2\pi\alpha(1-\beta)\dot{M}_2a^2/P$,
where  $a$ is the orbital separation and $P$ the orbital period, we can define $\alpha$ as the ratio $(d/a)^2$,
where $d$ is the distance from the center of mass (CM) of the binary system from which the matter leaves. 
Finally, the function $\Lambda(q,\alpha,\beta)$ is defined as
\begin{equation} 
\begin{split}
\Lambda(q,\alpha,\beta) 
= 1-\frac{3}{2}\beta q - \frac{1-\beta}{2} \frac{q}{1+q} -1.5\; \alpha (1-\beta)(1+q).
\end{split}
\end{equation}

\cite{iaria_15} assumed that a large fraction of the 
mass transferred by the CS leaves the system at the inner Lagrangian point.
Since the masses of the two bodies are unknown,  we 
estimate    the ratio $b_1/a$ as a function of the mass ratio $q$ between 0.005 and 0.1,
where $b_1$ is the distance of the inner Lagrangian point from the NS.   We note that 
\cite{Plavec_64} found an expression of $b_1/a$ for values of $q$ higher than 0.1 and we verified {a posteriori} that 
it does not work for the range of $q$ of    interest to us.   We derived a more generic expression, 
\begin{equation} 
\frac{b_1}{a}=0.702-0.948\;q+ 2.77\;q^2-0.0825 \log(q),
\label{eq:L1}
\end{equation} 
imposing that the sum of the gravitational forces of the two bodies and the centrifugal force is null at the inner Lagrangian point (L$_1$); the accuracy of this relation is 0.5\%   
for $q$ between 0.005 and 0.1.  As   reported above the free parameter $\alpha$ is defined as $(d/a)^2$, where $d$ indicates the distance to the CM from where the binary system loses specific angular momentum due to mass loss. The minimum value of  $d/a$
corresponds to the ejection point coinciding with  the inner Lagrangian point (i.e.,  
$d/a=b_1/a-q/(1+q)$, which gives 0.726 for $q=0.048$), while it is reasonable  to assume that the maximum value of $d/a$ corresponds to the ejection point coinciding with  the CS position (i.e.,    $d/a=1-q/(1+q)$, which gives 0.954 for $q=0.048$). Consequently,  in the following we explore the range of  $d/a$ between 0.726 and 0.954.

Since we expect $q<0.1$ we can approximate  the term $q/(1+q)$ as $\simeq q$  in  eq. \ref{eq:m2}.  By differentiating   eq. \ref{eq:m2} and some algebraic manipulation
we obtain
\begin{equation} 
\begin{split}
\frac{\dot{m}_2}{m_2}
= 
-\frac{\dot{P}}{P} +\frac{3}{2}\frac{\dot{b_f}}{b_f}+\\ +\frac{1}{2} \frac{0.6q^{2/3}+q^{1/3}(1+q^{1/3})^{-1}-\ln(1+q^{1/3})}{0.6q^{2/3}+\ln(1+q^{1/3})} (1+\beta q) \frac{\dot{m}_2}{m_2}.
\end{split}
\label{eq:m2dotonm2}
\end{equation} 
As suggested by \cite{Rappaport_87}, the term $\dot{b}_f/b_f$ is much smaller than $\dot{m}_2/m_2$; neglecting 
that term we can rewrite eq. \ref{eq:m2dotonm2} in the compact form
\begin{equation} 
\begin{split}
\frac{\dot{m}_2}{m_2}
= 
-\frac{\dot{P}}{P} \left[ 1-\frac{1}{2} \Gamma(\beta,q) \right]^{-1}, 
\end{split}
\label{eq:m2_seconda}
\end{equation} 
 where
 \begin{equation*} 
 \Gamma(\beta,q)= 
 \frac{0.6q^{2/3}+q^{1/3}(1+q^{1/3})^{-1}-\ln(1+q^{1/3})}{0.6q^{2/3}+\ln(1+q^{1/3})} (1+\beta q).
 \end{equation*} 
 Combining eqs. \ref{eq:m2dotonm2_1} and \ref{eq:m2_seconda}
 we obtain
 \begin{equation} 
\begin{split}
\dot{P}_{-11}=0.218 \frac{1-\frac{1}{2} \Gamma(\beta,q)}{\Lambda(q,\alpha,\beta)} m_1^{5/3} q (1+q)^{-1/3}P_h^{-5/3} {\rm \; s \;s^{-1}},
\end{split}
\label{eq:pdot}
\end{equation} 
where $\alpha$ only depends   on $q$ and
$\dot{P}_{-11}$ is the orbital period derivative in units 
of $10^{-11}$. 

On the other hand, the source luminosity 
can be written as $L_x = -\beta (G M_1 \dot{M}_2)/R_{NS}$,
where $R_{NS}$ is the NS radius. This relation can be opportunely rewritten as 
$L_x \simeq 2.64\times10^{53} \beta m_1^2 q (\dot{m}_2/m_2)$ erg s$^{-1} $ for a NS radius of 10 km;  combined    with  eq. \ref{eq:m2_seconda} it becomes 
\begin{equation} 
\begin{split}
L_{37}=73.33 \; \beta \; m_1^2 q \left[ 1-\frac{1}{2} \Gamma(\beta,q) \right]^{-1} \frac{\dot{P}_{-11}}{P_h} {\rm \; erg \;s^{-1}},
\end{split}
\label{eq:lumin}
\end{equation} 
 where $L_{37}$ is the luminosity in units of $10^{37}$. 
 \begin{figure*}
\centering
\includegraphics[scale=.344]{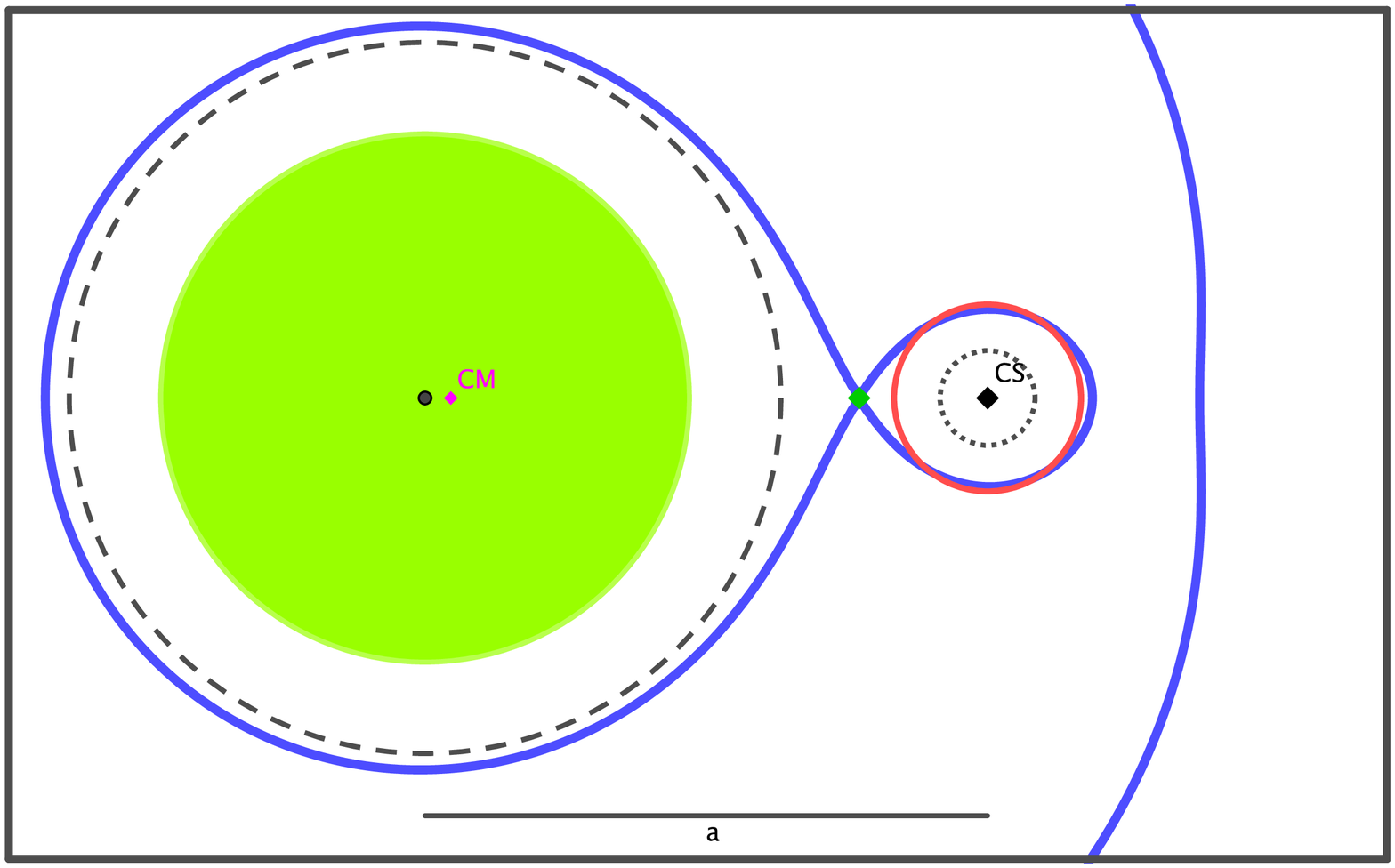}
\includegraphics[scale=.376]{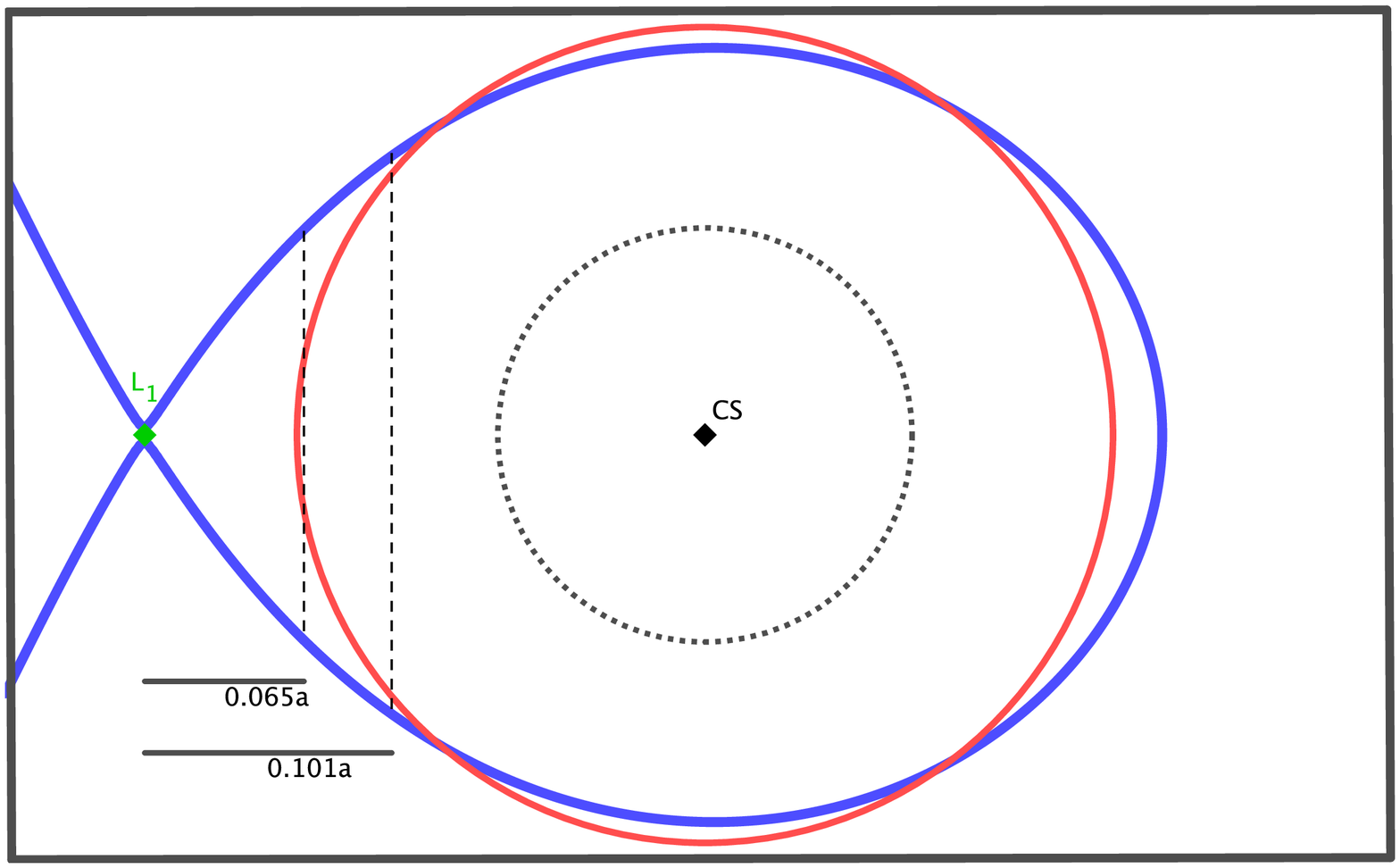}\\
\caption{  Top view in scale  of XB 1916-053 for a mass ratio $q=0.048$. The left and right panel shows the whole binary system and a zoom of it close to the inner Lagrangian point, respectively.  The blue curve indicates the Roche lobes of the two stars. The dashed and red circle  represent the NS Roche-lobe radius   and CS Roche-lobe radius. The magenta, black, and green diamond points are the position of the center of mass (CM),  the companion star, and the inner Lagrangian point $L_1$. The dotted circle is the radius of the companion star, without thermal inflation.  The green circle represents the accretion disk truncated at $0.47 a$ (see  text). The two black dashed segments indicate  the distances from $L_1$ at which the matter leaves the  binary system for a NS mass of 1.4 M$_{\odot}$ at a luminosity of $0.41 \times 10^{37}$ erg s$^{-1}$ and $1.46 \times 10^{37}$ erg s$^{-1}$ going from the closest to the farthest.} 
\label{figure:9b}
\end{figure*} 

From the study of the dip arrival times 
we infer an orbital period derivative   
$\dot{P} = (1.46 \pm 0.03) \times 10^{-11}$ s s$^{-1}$ and an 
orbital period   $P=3000.662(1)$ s. Studying    eq. \ref{eq:pdot} for $q=0.048$,
we explored   the pairs of 
$\beta$ and $d/a$ for which it is possible to obtain the $\dot{P}$  value  assuming 
    NS masses of 1.4, 1.6, and 1.8 M$_{\odot}$.

  To explain a $\dot{P}$  value of    $(1.46 \pm 0.03) \times 10^{-11}$ s s$^{-1}$,
 a non-conservative mass transfer  is required; we obtain that the parameter $\beta$ has to be smaller 
than 0.35 for each value of the NS mass studied. 
We find that the closer the ejection point is to the Lagrangian point, the smaller the fraction of accreted matter is.
The possible pairs of $\beta$ and $d/a$  are indicated with the red curve in Fig. \ref{figure:9}.

A further constraint on the ejection point and on the fraction of matter accreting onto the NS 
is given by the luminosity determined     in eq. \ref{eq:lumin}. 
 The distance of the source can be estimated using the flux during 
the type I X-ray bursts observed using {\it Rossi-XTE} data, as done by \cite{Galloway_08}. 
Using  eq. 10 in \cite{iaria_15}, given that the flux measured during 
the bursts showing photospheric radius expansion (PRE)
is $(2.9 \pm 0.4) \times 10^{-8}$ erg s$^{-1}$ cm$^{-2}$   and  
that  the NS  photospheric radius  is $r_{PRE}=1.1$  in units of 10 km  for XB 1916-053, we find 
that the distance $D$ to the source is 
$8.9\pm0.6$ kpc, $9.3\pm0.6$ kpc, and $9.6\pm0.7$ kpc for a NS mass of 1.4, 1.6, and 1.8 M$_{\odot}$, respectively. 
Adopting the 0.1-100 keV  unabsorbed flux obtained from spectra A, B, C, and D, 
we infer that  the luminosity varies between $0.41 \times  10^{37}$ and $1.46 \times 10^{37}$ erg s$^{-1}$ for a NS mass of 1.4 M$_{\odot}$, between 
$0.45 \times  10^{37}$ and $1.6 \times 10^{37}$ erg s$^{-1}$ for a NS mass of 1.6 M$_{\odot}$, and  between 
$0.48 \times  10^{37}$ and $1.7 \times 10^{37}$ erg s$^{-1}$ for a NS mass of 1.8 M$_{\odot}$ by taking into account the error associated with the distance and considering a 10\%  relative error for the  flux.  A similar ranges of luminosities
were also observed by \cite{Boirin_00}, which confirmed that the source belongs to the Atoll class. 
The 0.1-100 keV unabsorbed luminosities with the corresponding uncertainties are shown in Table \ref{tab:diagnostic_abcd}
for spectra A, B, C, and D.

We plotted the boundary values of luminosity for $q=0.048$ in Fig. \ref{figure:9}; since the luminosity   depends on $\beta$ but not on the quantity $\alpha=(d/a)^2$ (see eq. \ref{eq:lumin}), it is represented with horizontal lines. 
 The observed luminosities  constrain $\beta$ 
to be lower than 15\%, independently of the assumed NS mass. The variation in observed luminosity can be explained 
 with a change of the ejection point from which the matter leaves the binary system. 
 At low luminosity the matter leaves the system close to the inner Lagrangian point, and the fraction of mass accretion rate is 
 only 3\%, while at higher luminosity the matter leaves the system from a larger  distance and $\beta=0.11$. 
 We show a sketch of the top-view geometry (in scale) of XB 1916-053 assuming 
 $q=0.048$ and a NS mass of 1.4 M$_{\odot}$ in  Fig. \ref{figure:9b}. The left panel shows the whole binary system, while the right panel shows  the companion star and the region close the inner Lagrangian point. 
  The two black dashed segments indicate the boundaries of the ejection region measured from  $L_1$, the  segment distant 0.065$a$ from  $L_1$. The lower boundary  corresponds to a luminosity of $0.41 \times 10^{37}$ erg s$^{-1}$, while the upper boundary (0.101$a$) corresponds to a luminosity of  $1.46 \times 10^{37}$ erg s$^{-1}$. Similar results are obtained assuming a NS mass of 1.6 and 1.8  M$_{\odot}$.    
Hence, we have shown that abandoning the stringent hypothesis that the matter could leave the system only from the inner Lagrangian point, as supposed by \cite{iaria_15},     a NS mass of 1.4  M$_{\odot}$ also  describes the observed scenario well.

We  note that this scenario  
assumes that the observed luminosity $L_{\rm obs}$ is that emitted from the source   $L_{\rm real}$; however,  XB 1916-053 is a dipping source, and so we expect an inclination angle larger than $60^{\circ}$. Since we do not observe eclipses, we can estimate the upper limit on the inclination angle $i$ from the relation $i=90^{\circ}-\arctan(R_{L_2}/a)$, where $R_{L_2}$ is the Roche lobe radius of the CS,  finding   that $i<80^{\circ}$. If the 
emission is not spherical from the innermost region of the system then $L_{\rm obs} = L_{\rm real}\cos{i}$ and we should take into account this amplification factor. By
assuming the arbitrary value of $i=70^{\circ}$, the real luminosity varies between  $1.2\times 10^{37}$ erg s$^{-1}$  and $4.3 \times 10^{37}$ 
 erg s$^{-1}$. These two values of luminosity combined with the necessity to reproduce 
 an   orbital period derivative of $1.46 \times 10^{-11}$ s s$^{-1}$ gives a value of  $\beta$ between 0.08 and 0.34. In this scenario the matter leaves the system 
 at a distance from $L_1$ close the CS position.

 We note that the mass transfer rate is close to the Eddington limit in this evolutive stage of the source. However, the secular evolution  predicts that this system should have a mass transfer rate that is  two orders of magnitude lower \cite[see, e.g.,][]{Heinke_13} for a conservative mass transfer scenario. 
 \begin{table} 
\centering
\scriptsize 
\begin{threeparttable}
\caption{Parameters of XB 1916-053}
\begin{tabular}{l@{\hspace{2pt}}c@{\hspace{\tabcolsep}}}
\hline

 \hline 
Parameter &      \\      
     \hline 
\hline

Mass ratio $q$ & 0.048  \\\\  
 
Orbital period $P$ (s) & 3000.66    \\
Orb. period derivative $\dot{P}$ (s s$^{-1}$) & $(1.46 \pm 0.03) \times 10^{-11}$\\\\

Orbital separation $a${\textit{$\rm ^a$}}  (cm) & $3.54 \times 10^{10}$\\
Roche lobe radius of the CS, $R_{L_2}$ & $0.156a$\\
Roche lobe radius of the NS, $R_{L_1}$ & $0.594a$\\
Outer radius of the accretion disc, $r$ & $0.47a$ \\
Distance of the CM from the NS &  $0.046a$\\
Distance of $L_1$ from the NS, $b_1$ & $0.772a$ \\ 
Inclination angle $i$  & $60^{\circ}<i<80^{\circ}$\\
Bloating thermal factor $b_f$ of the CS{\textit{$\rm ^a$}}  & 1.92 \\
Observed luminosity{\textit{$\rm ^a$}}  ($10^{37}$ erg s$^{-1}$) & 0.41-1.46\\
 $\beta${\textit{$\rm ^b$}}  & 0.03-0.11\\
 Boundaries of the ejection region{\textit{$\rm ^c$}}&0.065$a$-0.101$a$ \\

\hline
\hline
\end{tabular}
     \begin{tablenotes}
\item[a] Assuming a NS mass of 1.4 M$_{\odot}$.
\item[b] Fraction of the mass  transferred  
from the CS that accretes onto the NS.
\item[c] The boundaries are measured from the inner Lagrangian point $L_1$.
 \end{tablenotes}
\label{tab:resume}
 \end{threeparttable}
\end{table}  
 Finally, \cite{Lasota_08} discussed the stability of the helium accretion disk in ultracompact X-ray  binary systems. From their estimations, the authors suggested that XB 1916-053 should be transient; however, the source is observed as persistent. The authors 
 showed three possible explanations of that incongruity: i) the adopted mass accretion rates (bolometric luminosities) could be underestimated, ii)  the outer disk radius is overestimated, iii)  the CS is not a  pure-helium star,  but still contains some   hydrogen. From our analysis all the caveats highlighted by the authors are verified.  \cite{Lasota_08} assumed a mass accretion rate of $7.6 \times 10^{-10}$ M$_{\odot}$ yr$^{-1}$, while  we find  from the estimated bolometric fluxes that 
 the mass accretion rate is between $3 \times 10^{-10}$ and $1.2 \times 10^{-9}$ M$_{\odot}$ yr$^{-1}$ for a NS mass of 1.4 M$_{\odot}$  and NS radius of 10 km. We show that the outer radius of the accretion disk is not truncated at $r_T=R_T/a=0.6/(1+q),$ which corresponds to $R_T=0.57\;a$ for $q=0.048$;   it is smaller because of the 3:1 resonance, and we find 
 $r = 0.47\; a$. Finally, the fraction of hydrogen is close to 0.2 for a NS mass of 1.4  M$_{\odot}$, and  increases for higher values of NS mass.   We summarized the main parameters of the components of XB 1916-053 in Table \ref{tab:resume}.

In the following we arbitrarily assume a NS mass of 1.4 M$_{\odot}$. Using the mass ratio      $q=0.048$, the   CS mass 
is 0.07  M$_{\odot}$, the   bloating thermal factor of the CS is $b_f=1.92$, and the fraction of hydrogen  of the  CS is close to 20\%.  
We find that the orbital separation of the binary system is $a\simeq 3.54 \times 10^{10}$ cm, while the outer radius of the accretion disk is $r_{\rm disk}= 1.7 \times 10^{10}$ cm.

\cite{iaria_15} suggest that  the sinusoidal modulation obtained from the LQS ephemeris could due to be the presence of a third body gravitationally bound to the X-ray binary system. Assuming the existence of a third body of mass $M_3$, the binary system   orbits around the new CM  of the triple system.
The distance of the binary system from the new CM is given by $a_x = a_{\rm{bin}} \sin i = A\; c$,   where $i$ is  the inclination of the binary plane to the plane of the sky, $A$ is the amplitude of the sinusoidal function obtained from the   ephemeris  shown in Eq. \ref{linear_quad_sin_eph}, and $c$ is the speed of the light. 
We obtain $a_x = (3.9\pm0.4) \times 10^{12}$ cm. To infer the mass of the  third body we can write its mass function 
as
\begin{equation*}
\frac{M_3 \sin i}{(M_3+M_{\rm{bin}})^{2/3}}=\left(\frac{4\pi^2}{G}\right)^{1/3} \frac{a_x}{P_{\rm{mod}}^{2/3}},   
\end{equation*}
where $M_{\rm{bin}}$ is the mass of the binary system.  Assuming arbitrarily $i = 70^{\circ}$, we found M$_3 \simeq 45$ M$_J$ under the hypothesis that the orbit of the third body is coplanar to the binary system.  

\begin{figure*}
\centering
\includegraphics[scale=.51]{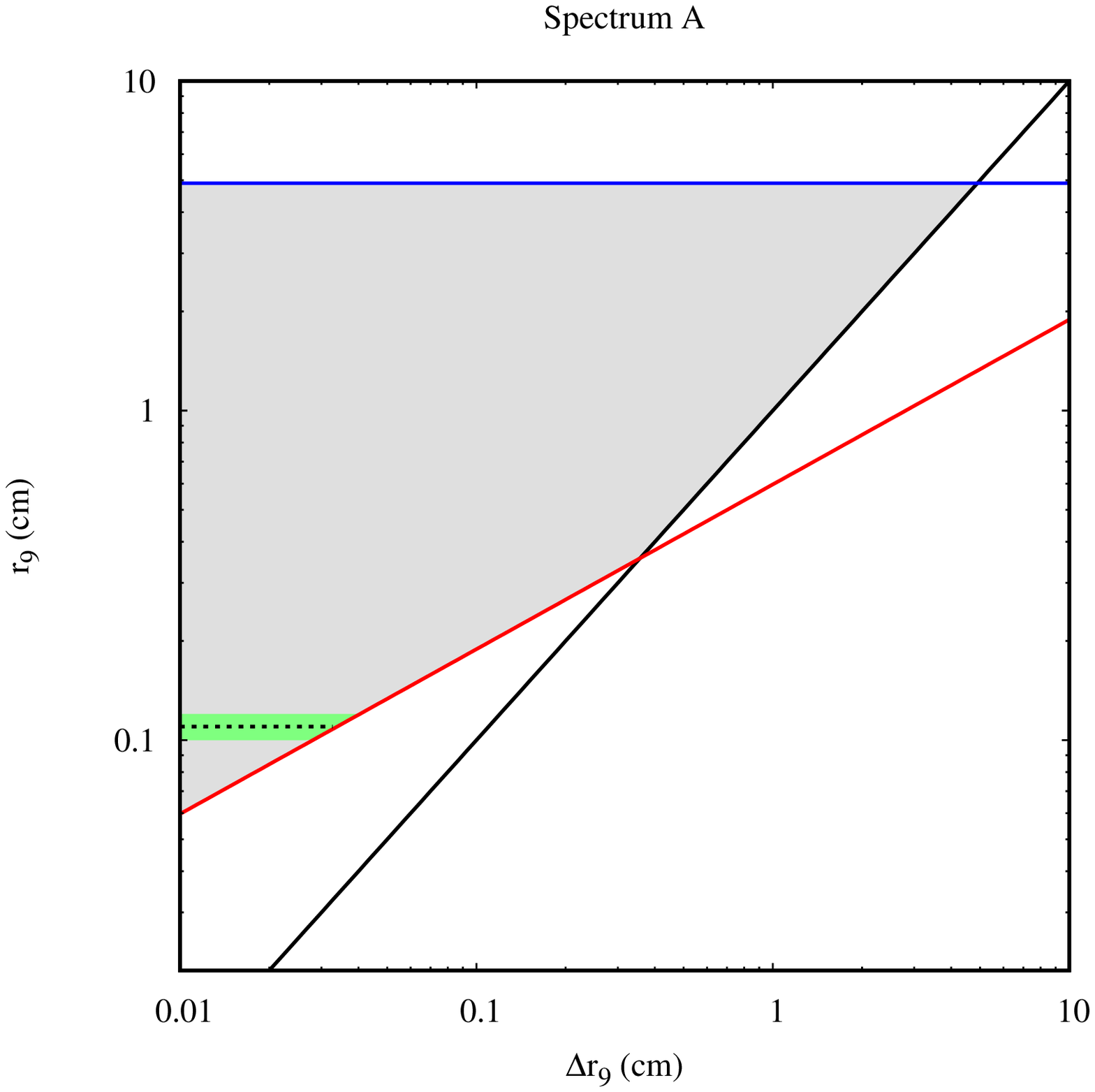}
\includegraphics[scale=.51]{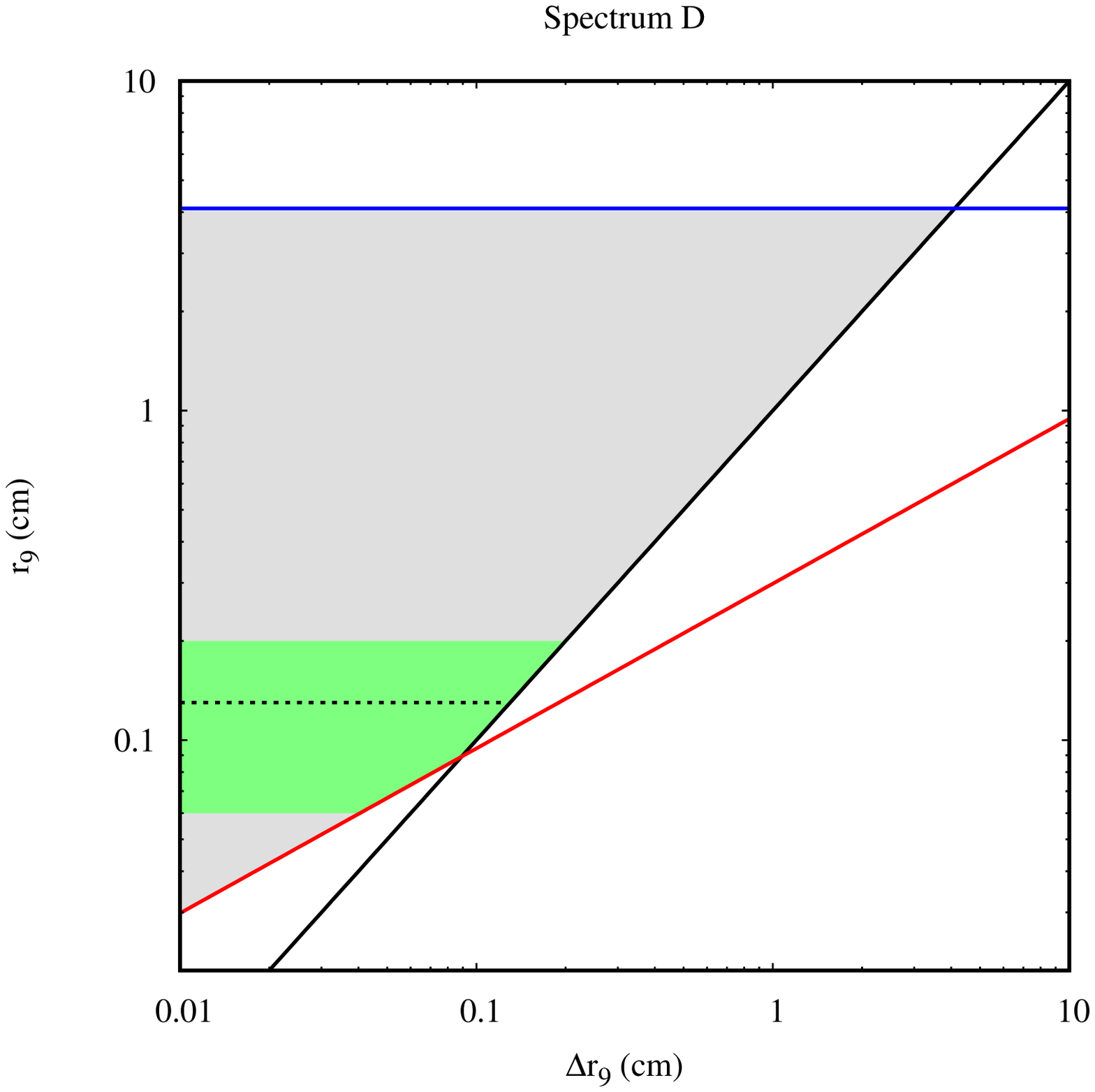} 
\caption{Distance $r$ of the absorber, in units of $10^9$ cm, from the NS with respect to  its thickness $\Delta r$ (in units  of $10^9$ cm). The gray area indicates the possible pairs of $r$ and $\Delta r$. 
The horizontal dashed line indicates the best-fit value of $r$ obtained assuming that the observed redshift is due to gravitational redshift for a NS mass of 1.4 M$_{\odot}$; the green area takes into account  the associated errors. The black, red, and blue lines are the constraint conditions (see text).}
\label{figure:10}
\end{figure*}

\subsection{Ionized absorber}

From the spectroscopic analysis  we 
investigated the nature of the ionized absorber around the compact object. 
\begin{table} 
\centering
\begin{threeparttable}
\caption{Plasma diagnostic of the spectra A, B, C, and D.}
\scriptsize
\begin{tabular}{l@{\hspace{\tabcolsep}}c@{\hspace{\tabcolsep}}c@{\hspace{\tabcolsep}}c@{\hspace{\tabcolsep}}c@{\hspace{\tabcolsep}}}
\hline
\hline
 &     Spectrum A & Spectrum B& Spectrum C & Spectrum D  \\

\hline

 N$_{\ion{Ne}{x}}${\textit{$\rm ^a$}}  &  
$0.39 \pm 0.09$ & 
 $0.37 \pm 0.11$  &
$0.31 \pm 0.11$ &
-- \\

N$_{\ion{Mg}{xii}}$   &    
$0.28 \pm 0.04$  &  
$0.24 \pm 0.04$&
$0.26\pm0.04$&
$0.24\pm0.11$\\           

N$_{\ion{Si}{xiv}}$   &    

$0.70 \pm 0.03$  &  
$0.59 \pm 0.04$&
$0.72\pm0.07$&
$1.5\pm0.4$\\        

N$_{\ion{S}{xvi}}$   &    

$0.74 \pm 0.15$  &  
$0.61 \pm 0.15$&
$1.0\pm0.2$&
--\\    

N$_{\ion{Ca}{xx}}$   &    

$1.0 \pm 0.3$  &  
--&
$0.8\pm0.3$&
--\\    

N$_{\ion{Fe}{xxv}}$   &    

$1.5 \pm 0.3$  &  
--&
--&
--\\    

N$_{\ion{Fe}{xxvi}}$   &    

$7.2 \pm 1.3$  &  
$9.6 \pm 1.1$&
$8.5\pm1.5$&
--\\\\    

L$_x${\textit{$\rm ^b$}}  & 
$1.5\pm0.2$ &
$1.0 \pm0.2$ &
$0.67\pm0.11$ &
$0.41\pm0.06$ \\\\

[N$_{\ion{Ne}{x}}$]/[N$_{\rm Ne}$]{\textit{$\rm ^c$}}   &  
$0.24\pm0.08$ & 
 $0.31\pm0.10$  &
$0.23\pm0.09$ &
-- \\

[N$_{\ion{Mg}{xii}}$]/[N$_{\rm Mg}$]   &    
$0.37\pm0.09$   &  
$0.43\pm0.09 $&
$0.41\pm0.10$ &
$0.7\pm0.4$\\           

[N$_{\ion{Si}{xiv}}$]/[N$_{\rm Si}$]   &    

$1.1\pm0.2$  &  
$1.3\pm0.2 $&
$1.4\pm0.3$&
$5\pm3$\\        

[N$_{\ion{S}{xvi}}$]/[N$_{\rm S}$]   &    

$3.0\pm0.9$  &  
$3.3\pm0.9$&
$4.7\pm1.2$&
--\\    

[N$_{\ion{Ca}{xx}}$]/[N$_{\rm Ca}$]   &    

$24\pm9$ &  
--&
$23\pm10$&
--\\    

[N$_{\ion{Fe}{xxv}}$]/[N$_{\rm Fe}$]    &    

$2.5\pm0.9$ &  
--&
--&
--\\    

[N$_{\ion{Fe}{xxvi}}$]/[N$_{\rm Fe}$]    &    

$12\pm3$  &  
$22\pm4$&
$17\pm4$&
--\\ \\

\hline
\hline
\end{tabular}
      \begin{tablenotes}
 \item[a] The equivalent column density of the ions 
 shown  are in units of 10$^{17}$ atoms cm$^{-2}$.
\item[b] Unabsorbed extrapolated luminosity in the 0.1-100 keV energy range. The luminosity is in units of $10^{37}$ erg s$^{-1}$ for a NS mass of 1.4 M$_{\odot}$.
\item[c] The ratio is in units of $10^{-2}$. 
 \end{tablenotes}
\label{tab:diagnostic_abcd}
 \end{threeparttable}
\end{table}  
Initially we estimated the equivalent column densities 
of the ions producing the absorption lines in  
spectra A, B, C, and D.  To do this we used the relation shown by \cite{Spitzer_78}
\begin{equation*}
\frac{W_\lambda}{\lambda}=\frac{\pi e^2}{m_ec^2} N_j \lambda f_{ij} \simeq 8.85 \;10^{-13}    N_j  \lambda f_{ij}, 
\end{equation*}
where $ N_j$ is the equivalent column density for the relevant species, 
$f_{ij}$ is the oscillator strength, $W_\lambda$ is the equivalent width of the
line,  $\lambda$ is the wavelength in centimeters, 
$e$ is the electron charge, $m_e$ is the electron mass, and  $c$ is the speed of the light. A similar analysis was done for the dipping source XB 1254-690 \citep{Iaria_07} 
and X 1624-690 \citep{Iaria_07_1624} and, recently, for XB 1916-053 using {\it Suzaku} data  \citep{Gambino_19}. 
To infer the column densities 
of the ions  we assumed  $f_{ij} =0.416$  for all the 
H-like ions and $f_{ij} =0.798$ for the \ion{Fe}{xxv} ion
\citep{Verner_96_fij}, and
we used the equivalent widths shown in Table
\ref{tab:line}; the  inferred $ N_j$   values are shown in Table
\ref{tab:diagnostic_abcd}. 
We find that the   equivalent ion column  density is in the range between $10^{16}$ and $(7.2-8.5) \times 10^{17}$ cm$^{-2}$. 

We  estimated  the abundances of each ion with respect to the neutral element $N_{\rm ion}/N_{\rm el}$. To  this end,  we  used the values of the  neutral hydrogen equivalent column density of the ionized absorber, N$_{{\rm H}_{\xi}}$, shown in Table \ref{tab:mod1mod2} and we adopted the relation 
$N_{{\rm H}_{\xi}}=  N_{{\rm H}_{\xi}}/N_{\rm el}\; N_{\rm el}/N_{\rm ion} \;N_{\rm ion} $, where $N_{{\rm H}_{\xi}}/N_{\rm el}$ is the inverse of  a given element with respect to the hydrogen and $N_{\rm ion}$ is the value 
of the equivalent column density associated 
with  the ions shown in Table \ref{tab:diagnostic_abcd}. To infer the value of  $N_{\rm ion}/N_{\rm el}$ we adopted  the 
solar abundance shown by \cite{asplund_09}.  The values of $N_{\rm ion}/N_{\rm el}$  for spectra A, B, C, and D are shown in Table 
\ref{tab:diagnostic_abcd}.  We find that the ratio $N_{\rm ion}/N_{\rm el}$ is 
0.2\%, 0.4\%, 1-2\%, 3-4\%, $(24\pm10)$\%, 2.5\%, and 10-20\% for  \ion{Ne}{x}, \ion{Mg}{xii}, \ion{Si}{xiv},  \ion{S}{xvi},  \ion{Ca}{xx},  \ion{Fe}{xxv,} and  \ion{Fe}{xxvi}, respectively. In a scenario where the NS mass is 1.4 M$_{\odot}$, the fraction of hydrogen of the CS is close to 20\%; because  $X \simeq 0.7$ for solar abundance, we should roughly multiply  the adopted  abundances by a factor
of three. Consequently, the ion population with respect to a given neutral element should be reduced by a factor of $1/3$.

To estimate the region where the ionized absorber is located,
we used the definition of $\xi$ in {\sc zxipcf} from which we get  $r^2=L_x/(n_H\xi)$, where  $L_x$ is the unabsorbed luminosity in the 0.1-100 keV energy range, $r$  is the distance of the absorber from the central source, $\xi$ is the ionization parameter obtained by our fits, and $n_H$ is the hydrogen atom   density  of the absorber.  By combining the last equation with   N$_{{\rm H}_{\xi}}=n_H\Delta r$, where $\Delta r$ is the geometrical thickness of the absorber along the line of sight and   N$_{{\rm H}_{\xi}}$
is the equivalent hydrogen column density associated with the absorber, and by
imposing   $r>\Delta r$,  we obtained an
upper limit on $r$ that is $r<L_x/(N_{{\rm H}_{\xi}}\xi)$. Furthermore, we expect that the optical depth $\tau=\sigma_T N_{{\rm H}_{\xi}}$ 
of the absorber is lower than one, where $\sigma_T$ is the Thomson cross section. By rearranging $r^2=L_x/(n_H\xi)$, we obtain $r^2=L_x\sigma_T \Delta r/(\tau \xi),$ and consequently $r>(L_x \sigma_T \Delta r/\xi)^{1/2}$. By using the best-fit values of N$_{{\rm H}_{\xi}}$  and $\xi$ shown in Table \ref{tab:mod1mod2} and the 0.1-100 keV unabsorbed luminosities  shown in Table \ref{tab:diagnostic_abcd} we are able to constrain 
the region where the ionized absorber is placed. 

We show the distance of the absorber from the NS in units of $10^9$ cm with respect to the thickness $\Delta r$ in units of $10^9$ cm in Fig. \ref{figure:10}.
The condition $r<L_x/(N_{{\rm H}_{\xi}}\xi)$ is represented with a blue line,   $r>\Delta r$ is indicated with a black line, and $r>(L_x \sigma_T \Delta r/\xi)^{1/2}$ with a red line. The gray area gives the   pairs of $r$ and $\Delta r$  satisfying the conditions. Since the outer radius of the accretion disk is $1.7 \times 10^{10}$ cm and because   the maximum value of $r$ is roughly $5 \times 10^9$ cm, we conclude that the ionized absorber is placed in the innermost region of the system. 

After analyzing the same {\it Chandra} data, 
\cite{Trueba_20} suggests that the observed redshift in the absorption lines is probably due to   gravitational redshift, and   finds  the 
 distance of the absorber from the NS is $(2.7 \pm 1.7) \times 10^8$ cm and $(2.3 \pm 1.2) \times 10^8$ cm 
  for a NS mass of 1.4 M$_{\odot}$; these values   refer to spectra B and $\Gamma$ reported by the authors. By adopting the same scenario and using the best-fit values of the observed redshift shown in Table \ref{tab:mod1mod2} we find that the absorber is placed 
at $(1.1 \pm 0.1) \times 10^{8}$ cm, 
$(1.3 \pm 0.2) \times 10^{8}$ cm, $(1.2 \pm 0.2) \times 10^{8}$ cm, and $(1.3 \pm 0.7) \times 10^{8}$ cm for spectra A, B, C, and D, respectively. Our estimations are almost a factor of two smaller than the value estimated by  \cite{Trueba_20}, even if the values are compatible with each other. We show the best-fit value of $r$  and the associated errors with a dashed line and  the green area in Fig. \ref{figure:10}.  We find that the ionized absorber can reach the NS surface only in spectrum D when the source is dimmer; in the other cases the absorber forms a torus around the compact object. From N$_{{\rm H}_{\xi}}=n_H\Delta r$ we can infer the lower limit of $n_H$ assuming the maximum possible thickness of the absorber; 
we obtain $n_H>5 \times 10^{15}$ cm$^{-3}$, $n_H>2 \times 10^{15}$ cm$^{-3}$, $n_H>1.5 \times 10^{15}$ cm$^{-3}$, and $n_H>0.4 \times 10^{15}$ cm$^{-3}$ for spectra A, B, C, and D.

We repeated the same procedure to estimate  the distance of the absorber during the dip using the values in
Table \ref{tab:dip} and knowing that the 0.1-100 keV unabsorbed luminosity is $(0.6 \pm 0.1) \times 10^{37}$ erg s$^{-1}$. 
We find a weak constraint for $r$; it  
 is between  $2 \times 10^8$ and  $1.7 \times 10^{10}$ cm (the outer radius of the disk).  Because of the low 
 value of the ionization parameter  it is possible that the absorber is placed at the outer radius; 
 however, we cannot exclude   inner  radii.  We note that  the circularization radius  $r_{\rm circ} = 0.0859\; a\; q^{-0.426}$ \citep[valid for $q$ between 0.05 and 1;][]{Hessman_90} is $r_{\rm circ}\simeq 1.1\times 10^{10}$ cm for $q=0.048$, compatible with 
the obtained distances.

Finally, we explored the nature of the   \ion{O}{viii} absorption edge observed in the spectra during the persistent emission. 
Since its optical depth is $0.16$, assuming that $n_e\sim n_i$,  we can write $\tau = N_H \sigma_T$, where $\sigma_T$ is the Thomson cross section. We inferred an  equivalent hydrogen column density of $\sim 2.3 \times 10^{23}$ cm$^{-2}$. This value is compatible with the equivalent hydrogen column associated with the ionized absorber. 
We conclude that a small fraction of \ion{O}{viii}  could be present in the absorber. 

 \section{Conclusions}
 We updated the orbital ephemeris of the compact dipping source 
 XB 1916-053 by analyzing ten new {\it Chandra} observations and one {\it Swift/XRT} observation. 
 Three  new dip arrival times were extracted, allowing us to extend the available baseline from 36 to  40 years. Our analysis definitively excludes several models of ephemeris  suggested previously, leaving as a more likely candidate the quadratic orbital ephemeris including a large periodic modulation close to 25 years (LQS ephemeris in this work). 
 
 The quadratic term of the orbital ephemeris implies the 
 presence of an orbital period derivative of $1.46(3) \times 10^{-11}$ s s$^{-1}$, and allows us to refine the orbital period measurement to  3000.662(1) s. We confirm that such a fast orbital expansion  implies that mass transfer   should be highly non-conservative.
 Moreover, we find that for a NS mass of $1.4$ M$_{\odot}$ it is possible to obtain a compatible orbital period derivative and   luminosity values, releasing the stringent hypothesis that the mass leaves the system at the inner Lagrangian point. For a   NS mass of $1.4$ M$_{\odot}$ we find that the ejection point is distant from the inner Lagrangian point between $3 \times 10^8$ and $1.5 \times 10^9$ cm.  
 
 We show that the mass ratio of the system is $q=0.048$. This value is obtained   from the observed infrahump period, as already discussed in the literature by \cite{Hu_08}, and from the observed superhump period, which introduces the pressure term 
 due to a spiral wave to explain  the apsidal precession period of the accretion disk. 
 Our study allows us to explain the infrahump and superhump  periods  \citep{Chou_01}, assuming  that the accretion disk 
 has a prograde apsidal precession of 3.9 days and a retrograde nodal precession of 4.86 days. The outer radius of the 
accretion disk is truncated where the 3:1 resonance occurs (i.e.,  $1.7 \times 10^{10}$ cm for a NS mass of 1.4 M$_{\odot}$). 
 
 Assuming a NS mass of 1.4 M$_{\odot}$, we obtain that the CS mass is 0.07 M$_{\odot}$. The thermal bloating of the CS is 1.92 and its percentage of content  hydrogen is 20\%.

 We explain the periodic modulation observed in the orbital 
 ephemeris as the presence of a third body orbiting around the binary system. Assuming a co-planar orbit with respect to the binary system, an inclination angle of $70^{\circ}$, and a hierarchical triple system, we find that the mass of the third body is $M_3 \simeq 45$ Jovian masses.
 
 To perform our spectroscopic analysis we also included    an old Chandra observation, previously analyzed by 
 \cite{Iaria_06}. We observed a large variation in the count rate  for the different observations. Using 
 a Comptonized component for the continuum we estimated that the 0.1-100 keV unabsorbed luminosity varies   from $0.41 \times 10^{37}$ erg $^{-1}$ to $1.46 \times 10^{37}$ erg $^{-1}$ during the persistent emission. 
 
 The spectra show  the presence of prominent absorption lines associated with \ion{Ne}{x}, \ion{Mg}{xii}, \ion{Si}{xiv}, \ion{S}{xvi,} and \ion{Fe}{xxvi}.  To fit these lines 
 we included in the model the component  {\sc zxipcf} 
 that assumes the presence of an ionized absorber between the central object and the observer along the line of sight. 
 We find that the equivalent hydrogen column density  associated with the ionized absorber ranges from  $9 \times 10^{22}$ cm$^{-2}$ to $2 \times 10^{23}$ cm$^{-2}$, going from the dimmest to the brightest spectrum. The  ionization parameter is close to log${(\xi)}=4.3$. From our diagnostic, under the assumption that the observed redshift of the line produced in the absorber  is due to gravitational redshift, as recently discussed by \cite{Trueba_20}, 
 we find that the absorber has a hydrogen atom density $n_H$ higher than 
 $10^{15}$ cm$^{-3}$ and it is located far from the central source at $1 \times 10^8$ cm for a NS mass of 1.4 M$_{\odot}$. 
 
 We extracted the dip spectrum finding that the
 equivalent hydrogen column density  associated with the ionized absorber is  $6 \times 10^{23}$ cm$^{-2}$ and 
 the  ionization parameter is close to log${(\xi)}=2.8$. Finally, 
 we find that  during the dip   the ionized absorber is placed  at a greater distance, which    could be compatible with the edge of the 
 accretion disk ($1.7 \times 10^{10}$ cm) even if we cannot exclude that it is placed at the circularization radius  ($1.0 \times 10^{10}$ cm).

\section*{Acknowledgements}

This research has made use of data and/or software provided by the High Energy Astrophysics Science Archive Research Center (HEASARC), which is a service of the Astrophysics Science Division at NASA/GSFC and the High Energy Astrophysics Division of the Smithsonian Astrophysical Observatory.\\
The authors acknowledge financial contribution from the agreement
ASI-INAF n.2017-14-H.0, from INAF mainstream (PI: T. Belloni; PI: A. De Rosa)
and from the HERMES project financed by the Italian Space
Agency (ASI) Agreement n. 2016/13 U.O.
RI and TDS acknowledge the research
grant iPeska (PI: Andrea Possenti) funded under the INAF
national call Prin-SKA/CTA approved with the Presidential Decree
70/2016. 

\bibliographystyle{aa} 
\bibliography{biblio}
\end{document}